\documentclass[12pt, epsfig]{article}
\usepackage{epsfig}
\usepackage{amssymb}
\usepackage{amsfonts}

\begin{document}


\newcommand{\beq}{\begin{equation}}
\newcommand{\eeq}{\end{equation}}
\newcommand{\bea}{\begin{eqnarray}}
\newcommand{\eea}{\end{eqnarray}}
\newcommand{\beqn}{\begin{eqnarray}}
\newcommand{\eeqn}{\end{eqnarray}}
\newcommand{\beas}{\begin{eqnarray*}}
\newcommand{\eeas}{\end{eqnarray*}}
\newcommand{\defi}{\stackrel{\rm def}{=}}
\newcommand{\non}{\nonumber}
\newcommand{\bquo}{\begin{quote}}
\newcommand{\enqu}{\end{quote}}
\newcommand{\p}{\partial}


\def\de{\partial}
\def\Tr{ \hbox{\rm Tr}}
\def\const{\hbox {\rm const.}}
\def\o{\over}
\def\im{\hbox{\rm Im}}
\def\re{\hbox{\rm Re}}
\def\bra{\langle}\def\ket{\rangle}
\def\Arg{\hbox {\rm Arg}}
\def\Re{\hbox {\rm Re}}
\def\Im{\hbox {\rm Im}}
\def\diag{\hbox{\rm diag}}

\def\stroke{\vrule height8pt width0.4pt depth-0.1pt}
\def\topfleck{\vrule height8pt width0.5pt depth-5.9pt}
\def\botfleck{\vrule height2pt width0.5pt depth0.1pt}
\def\Zmath{\vcenter{\hbox{\numbers\rlap{\rlap{Z}\kern 0.8pt\topfleck}\kern
2.2pt\rlap Z\kern 6pt\botfleck\kern 1pt}}}
\def\Qmath{\vcenter{\hbox{\upright\rlap{\rlap{Q}\kern
3.8pt\stroke}\phantom{Q}}}}
\def\Nmath{\vcenter{\hbox{\upright\rlap{I}\kern 1.7pt N}}}
\def\Cmath{\vcenter{\hbox{\upright\rlap{\rlap{C}\kern
3.8pt\stroke}\phantom{C}}}}
\def\Rmath{\vcenter{\hbox{\upright\rlap{I}\kern 1.7pt R}}}
\def\Z{\ifmmode\Zmath\else$\Zmath$\fi}
\def\Q{\ifmmode\Qmath\else$\Qmath$\fi}
\def\N{\ifmmode\Nmath\else$\Nmath$\fi}
\def\C{\ifmmode\Cmath\else$\Cmath$\fi}
\def\R{\ifmmode\Rmath\else$\Rmath$\fi}


\def\QATOPD#1#2#3#4{{#3 \atopwithdelims#1#2 #4}}
\def\stackunder#1#2{\mathrel{\mathop{#2}\limits_{#1}}}
\def\stackreb#1#2{\mathrel{\mathop{#2}\limits_{#1}}}
\def\Tr{{\rm Tr}}
\def\res{{\rm res}}
\def\Bf#1{\mbox{\boldmath $#1$}}
\def\balpha{{\Bf\alpha}}
\def\bbeta{{\Bf\beta}}
\def\bgamma{{\Bf\gamma}}
\def\bnu{{\Bf\nu}}
\def\bmu{{\Bf\mu}}
\def\bphi{{\Bf\phi}}
\def\bPhi{{\Bf\Phi}}
\def\bomega{{\Bf\omega}}
\def\blambda{{\Bf\lambda}}
\def\brho{{\Bf\rho}}
\def\bsigma{{\bfit\sigma}}
\def\bxi{{\Bf\xi}}
\def\bbeta{{\Bf\eta}}
\def\d{\partial}
\def\der#1#2{\frac{\d{#1}}{\d{#2}}}
\def\Im{{\rm Im}}
\def\Re{{\rm Re}}
\def\rank{{\rm rank}}
\def\diag{{\rm diag}}
\def\2{{1\over 2}}
\def\ntwo{${\cal N}=2\;$}
\def\4N{${\cal N}=4$}
\def\none{${\cal N}=1\;$}
\def\x{\stackrel{\otimes}{,}}
\def\beq{\begin{equation}}
\def\eeq{\end{equation}}
\def\ba{\beq\new\begin{array}{c}}
\def\ea{\end{array}\eeq}
\def\be{\ba}
\def\ee{\ea}
\def\stackreb#1#2{\mathrel{\mathop{#2}\limits_{#1}}}

\def\changed#1{{\bf #1}}

\def\baselinestretch{1.0}

\begin{titlepage}

\begin{flushright}
IFUP-TH/2007-23 \\
\end{flushright}
\vspace{3mm} 

\begin{center}
{\Large \bf  Type I Non-Abelian Superconductors
\\[2mm] in Supersymmetric Gauge Theories }
\end{center}
\vspace{5mm} 

\begin{center}
{\bf Roberto~Auzzi$^{a}$, \bf Minoru~Eto$^{a,b}$ and Walter~Vinci$^{a,b}$}
\end {center}
\begin{center}
\vskip 1.5em $^a$ {\it Department of Physics, University of Pisa Largo Pontecorvo, 3,   Ed. C,  56127 Pisa, Italy }
\\ $^b$ {\it INFN, Sezione di Pisa, Largo Pontecorvo, 3, Ed. C, 56127 Pisa, Italy } \vspace{12mm}
\end{center}

\vspace{10mm}

\begin{abstract}
Non-BPS non-Abelian vortices with $\mathbb{CP}^1$ internal moduli space are studied in an ${\cal N}=2$ supersymmetric
$U(1)\times SU(2)$ gauge theory with adjoint mass terms. For generic internal orientations the classical force between
two vortices can be attractive or repulsive. On the other hand, the mass of the scalars in the theory is always less
than that of the vector bosons; also, the force between two vortices with the same $\mathbb{CP}^1$ orientation is
always attractive: for these reasons we interpret our model as a non-Abelian generalization of type I superconductors.
We compute the effective potential in the limit of two well separated vortices. It is a function of the distance and of
the relative colour-flavour orientation of the two vortices; in this limit we find an effective description in terms of
two interacting $\mathbb{CP}^1$ sigma models. In the limit of two coincident vortices we find two different solutions
with the same topological winding and, for generic values of the parameters, different tensions. One of the two
solutions is described by a $\mathbb{CP}^1$ effective sigma model, while the other is just an Abelian vortex without
internal degrees of freedom. For generic values of the parameters, one of the two solutions is metastable, while there
are evidences that the other one is truly stable.
\end{abstract}

\end{titlepage}

\section{Introduction}

According to the ideas of Mandelstam and 't Hooft \cite{TM}, confinement in non-Abelian gauge theories is due to a dual
Meissner effect. The electric flux between two heavy electric sources is confined by a dual Abrikosov-Nielsen-Olesen
vortex \cite{ANO}; the vortex has a constant energy per unit length (tension). This leads to a linear potential between
the probe charge and anti-charge. Due to the difficulties in analyzing strongly interacting non-Abelian gauge theories,
this picture  remained just a nice qualitative scheme for years, which could not be justified from first principles.

A breakthrough in this context was the Seiberg-Witten solution \cite{SW} of $\mathcal{N}=2$ super Yang-Mills theories;
they found massless monopoles at strong coupling. Adding a small $\mathcal{N}=2$ breaking mass term for the adjoint
field, the monopoles condense creating dual vortex strings which carry a chromoelectric flux. The details of
confinement in the Seiberg-Witten scenario are indeed quite different from QCD. The $SU(N_c)$ gauge symmetry is
spontaneously broken broken to $U(1)^{N_c-1}$ by the expectation value of an adjoint field and  the strings of the
theory carry an Abelian $U(1)^{N_c-1}$ charge. A careful examination shows that the ``hadronic'' spectrum is much
richer than that of QCD \cite{DSHSZ}.

Thus it is interesting to study some alternatives  to the Abelian Meissner effect, with the aim to find some close
relatives of QCD. The non-Abelian vortex discussed in Refs.~\cite{ht1,nab} is an interesting possibility in this
direction.
 This solitonic object has first
been studied in an $\mathcal{N}=2$ $U(N_c)$ gauge theory with $N_f=N_c=N$ quark hypermultiplets and with a
Fayet-Iliopolous term in order to keep the theory in the Higgs phase. The squark fields condense and break the gauge
symmetry; on the other hand the colour-flavour locked global symmetry is unbroken in the vacuum. The theory has vortex
solitons which spontaneously break this $SU(N)$ symmetry to $SU(N-1)\times U(1)$; due to the zero modes corresponding
to these broken symmetries, the moduli space is given by the quotient:
\[ \mathbb{CP}^{N-1}=\frac{SU(N)}{SU(N-1)\times U(1)}.\]
The classical moduli coordinate can be promoted to a field living on the vortex worldvolume; in this way vortex
solitons in a $3+1$ dimensional theory can be directly connected with a $\mathbb{CP}^{N-1}$ sigma model in $1+1$
dimension, which describes the macroscopic physics of the flux tube. Some reviews on this subject can be found in
Ref.~\cite{reviews};
 recent developments involve the Seiberg duality~\cite{seibvortex}, the Goddard-Nuits-Olive duality~\cite{Eto:2006dx},
generalizations to $SO(N)$ theories \cite{ferretti}, and possible phenomenological applications to high temperature
Yang-Mills and dense matter \cite{gorsky}.

The non-Abelian vortex can be studied in many different theoretical settings; indeed it is possible to start with an
$\mathcal{N}=1$ theory \cite{ssize,tongheterotic} or even with a non-supersymmetric theory \cite{gsy}. The details of
the effective $1+1$ dimensional sigma model are different due to  different number of fermions and various amount of
supersymmetry.
 Also the number of quantum vacua is different,
for example there are $N$ vacua in the $\mathcal{N}=2$ case~\cite{tong-cm,sy-cm,ht2}
 and just one vacuum in the
non-supersymmetric case~\cite{gsy}.

It is also interesting to study the non-Abelian vortex for higher winding numbers. In the $\mathcal{N}=2$ case, the
vortex is a BPS object with a big moduli space; as discussed in Ref. \cite{ht1,EINOS}, in the topological sector with
winding $k$, the dimension of the moduli space is $2 k N$. Some of these moduli correspond to the relative and global
positions of the component $k=1$ vortices; others to the global and relative orientations in the internal space. The
vortex solution and the moduli space for higher winding numbers has been discussed in
Refs.~\cite{hash,EINOS,composite,EKMNOVY,ehmno}.

In this paper, we study the impact on the vortices of the $\mathcal{N}=2$ model
 of some mass terms $\eta_0,\eta_3$
for the adjoint fields, which break the extended supersymmetry. For concreteness, we will discuss the case $
N_c=N_f=2$. The vortex with winding number one is not anymore a BPS object, but still has a $\mathbb{CP}^{1}$ moduli
space. On the other hand the physics for vortices with higher winding numbers is very different: almost all the flat
directions in the moduli space are lifted by the parameters $\eta_0,\eta_3$. The force between two vortices is not as
simple as in an Abelian superconductor, where we have attraction for type I vortices and repulsion for type II ones
\cite{Bo,jr}. There is a non-trivial dependence on the orientations of the two vortices
 in the internal space.

Even if the force between two vortices in our model is not attractive for all values of the vortex orientations
$\vec{n}_1,\vec{n}_2$, we have a close resemblance with type I Abelian vortices: we find that the scalars of the theory
are lighter than the vector bosons. Hence if we consider two well separated vortices, we have that the prevailing part
of the interaction is mediated by scalars and not by vectors. Moreover, for $\vec{n}_1=\vec{n}_2$ the force is always
attractive. We have also found evidences that the configurations which minimize the energy are always given by two
coincident vortices, just as in the type I Abelian case. For these reasons we call these objects "non-Abelian type I
vortices".

In Sect.~2 we describe the theoretical set-up. In Sect.~3 we discuss the non-BPS solutions for the vortices which live
in the Abelian subset of the theory, with emphasis on the sectors which have topological winding 1 and 2. In Sect.~4 we
study a more general configuration of two coincident vortices, both in the BPS case  and in the non-BPS case; a
potential for the vortex moduli space is found for $\eta_{0,3}\neq 0$. In Sect.~5 the interactions between two vortices
with a large separation distance are studied and the effective vortex potential is computed in this limit. In Sect.~6
the worldsheet description of the macroscopic physics is discussed, both for a single vortex and for two vortices at
large distance. In Sect.~7 we conclude the paper and make a general discussion. Some aspects specific to the large
$\eta_{0,3}$ limit are discussed in Appendix~A. In Appendix~B the BPS equations for two coincident vortices for
$\eta_0=\eta_3=0$ are provided.

After this work was finished, two papers about interactions of global non-Abelian vortices appeared \cite{verynewones};
however the details of these two models are quite different from the setting studied in this paper.

\section{Theoretical Set-Up}

\subsection{Lagrangian}

For the $U(1)$ gauge field $A_\mu^0$ and the $SU(2)$ gauge field $A_\mu^k$ ($k=1,2,3$)
the following conventions  are used:
\beq
A_\mu =
 \frac{\tau^k}{2} A_\mu^k +
\frac{1}{2}A^0_\mu,
\eeq
\[
\nabla_\mu =\partial_\mu-i \frac{\tau^k}{2} A_\mu^k -
\frac{i}{2} A^0_\mu, \qquad
 D_\mu a^k=\partial_\mu a^k + \epsilon^{klm} A_\mu ^l a^m.\]
The field strength is:
\beq F_{\mu \nu}=\partial_\mu A_\nu-\partial_\nu A_\mu-\frac{i}{4}
\left[ A_\mu^j \tau^j, A_\nu^k \tau^k\right],
\eeq
which in components (with the convention $F_{\mu \nu}=F_{\mu \nu}^k \tau^k/2$) reads:
\[  F_{\mu \nu}^i=\partial_\mu A_\nu^i-\partial_\nu A_\mu^i+
\epsilon^{ijk} A_\mu^j A_\nu^k.\]

We consider an $\mathcal{N}=2$ supersymmetric $U(1)\times SU(2)$ gauge theory with $N_f=2$ hypermultiplets together
with the following superpotential:
\beq W= \frac{1}{\sqrt{2}}
\left[ \tilde{Q}_f (a+a^k \tau^k+\sqrt{2} m_f) Q_f + W_0(a)+ W_3(a^k \tau^k)\right],\eeq where the terms
$W_{0,3}$ are of the form
\beq W_0=- \xi a + \eta_0 a^2, \,\,\, W_3(a^k \tau^k)=\eta_3 a^k a^k. \eeq
Here we have introduced two real positive mass parameters $\eta_0$ and $\eta_3$ for the adjoint scalars which break
${\cal N}=2$ SUSY to ${\cal N}=1$. $m_f$ is the mass of the hypermultiplets $Q_f,\tilde Q_f$ $(f=1,2)$ and $\xi$ is the
FI $F$-term parameter\footnote{A very similar Lagrangian was discussed in Refs. \cite{ssize}, \cite{tongheterotic}; in
that case the FI was in the D-term and not in the superpotential. This leads to different physics, the vortex is still
classically BPS saturated. For a discussion of the different settings that give BPS vortices, see
Refs.\cite{bpsornot}.}. Vortices in Abelian versions of this theoretical setting have been discussed in
Refs.~\cite{vy},\cite{abe}.

This kind of potential naturally arises from the $\mathcal{N}=2$, $SU(3)$ SQCD softly broken with a mass term of the
form $W=\eta \textrm{Tr} A^2$. Indeed, when the bare masses of the squarks are tuned to special values, there exist
true quantum vacua in which the non-Abelian gauge symmetry $SU(2)\times U(1)$ is preserved~\cite{nab}. The low energy
effective theory in these vacua is exactly the theory we are studying here\footnote{These quantum vacua exist only if
we have a sufficient number of flavours.  In this case semilocal vortices may be relevant~\cite{semilocal}. }.

 The bosonic part of the Lagrangian in Euclidean notation
 is (we use the same symbols for
the scalars as are used for the corresponding superfields):
\beqn
{\cal L} &=& \int d^4  x \bigg[ \frac{1}{4 e_3^2} |F_{\mu \nu}^k|^2 +\frac{1}{4 e_0^2} |F_{\mu \nu}|^2+
 \frac{1}{ e_3^2} |D_{\mu} a^k|^2+ \frac{1}{ e_0^2}  |\partial_{\mu} a|^2
 \nonumber\\[3mm]
 &+& \Tr (\nabla_\mu Q)^{\dagger} (\nabla_\mu Q)+
 \Tr (\nabla_\mu \tilde{Q}) (\nabla_\mu \tilde{Q}^{\dagger})+
 V(Q,\tilde{Q},a^k,a) \bigg],
 \label{azione-tutta}
 \eeqn
 where $e_0$ is the $U(1)$ gauge coupling and $e_3$ is $SU(2)$ gauge coupling.
 The potential $V$ is the sum of the following $D$ and $F$ terms:
 \beqn
V&=& \frac{e_3^2}{8}
 \left( \frac{2}{e_3^2} \epsilon^{ijk} \bar{a}^{j} a^k +
 \Tr (Q^\dagger \tau^i Q) -\Tr(\tilde{Q} \tau^i \tilde{Q}^\dagger) \right)^2
 \nonumber\\[3mm]
  & +&
 \frac{e_0^2}{8} \left(\Tr (Q^\dagger Q)-\Tr(\tilde{Q} \tilde{Q}^\dagger) \right)^2
 \nonumber\\[3mm]
  & +& \frac{e_3^2}{2} \left|\Tr (\tilde{Q} \tau^i Q)+2 \eta_3 a^i \right|^2 +
 \frac{e_0^2}{2} \left|\Tr (\tilde{Q}  Q)- \xi +2 \eta_0 a \right|^2
 \nonumber\\[3mm]
  & +& \frac{1}{2} \sum_{f=1}^2 |(a+\tau^i a^i+\sqrt{2} m_f) Q_f |^2+
 |(a+\tau^i a^i+\sqrt{2} m_f) \tilde{Q}^{\dagger}_{f} |^2 \,.
 \eeqn
The squark multiplets are kept massless in the remainder of the paper,
\[ m_f=0.\]
The vacuum of the theory which we are interested in is not changed by the parameters $\eta_{0,3}$:
\beq
Q=\tilde{Q}=\sqrt{ \frac{\xi}{2} }\left(\begin{array}{cc}
1 & 0\\
0 & 1 \\
\end{array}\right) ,\,\,\, a=0,\,\,\, a^b=0.
\label{eq:vac}
\eeq
For $\eta_0 \neq 0$, the theory has also another classical vacuum:
\beq Q=\tilde{Q}=0, \,\,\, a=\frac{\xi}{2 \eta_0}, \,\,\, a^k=0, \eeq
which ``runs away'' at infinity for $\eta_0=0$. In what follows, we consider the vacuum (\ref{eq:vac}) and the
non-Abelian vortices therein.

\subsection{Spectrum of the Theory}

The masses of the gauge bosons can easily be read of the Lagrangian:
\[ M_{U(1)}^2= \xi e_0^2, \,\,\, M_{SU(2)}^2= \xi e_3^2. \]
The masses of the scalars are given by the eigenvalues $M_i^2$ of the mass matrix (calculated in the vacuum
(\ref{eq:vac})):
\beq \label{mmassa}
\mathcal{M}=\frac{1}{2} \frac{\partial^2 V}{\partial s_i \partial s_j},\eeq where we denote by $s_k$
($k=1,2,\cdots,24$) the real scalar fields of the theory. The calculation is a bit tedious but quite straightforward (a
very similar situation is discussed in Ref. \cite{ssize}, in the case of a D-term Fayet-Iliopoulos). First of all,
there are four zero eigenvalues, which correspond to the scalar particles eaten by the Higgs mechanism. There is one
real scalar with mass $M_{S0}=M_{U(1)}$ which is in the same ${\cal N}=1$ multiplet as the $U(1)$ massive photon and
moreover there are also three scalars with a mass $M_{T0}=M_{SU(2)}$ in the same multiplet as the non-Abelian vector
field. The other mass eigenvalues are given by
\beq M_{S1,S2}^2= \xi e_0^2 + e_0^4 \eta_0^2\pm  \sqrt{2 \xi \eta_0^2 e_0^6+e_0^8 \eta_0^4} \ ;
\label{masse} \eeq
\[ M_{T1,T2}^2 =  \xi e_3^2 +  e_3^4 \eta_3^2 \pm  \sqrt{2 \xi \eta_3^2 e_3^6+e_3^8 \eta_3^4} \ , \]
where  the upper sign is for $M_{S1,T1}$ and the lower sign is for $M_{S2,T2}$. $M_{S1}$ and $M_{S2}$ have multiplicity
2; $M_{T1}$ and $M_{T2}$ have multiplicity 6. The mass of the fermions is obviously the same as the mass of the bosonic
degrees of freedom, because of the unbroken ${\cal N}=1$ supersymmetry.

Note that for $\eta_0=\eta_3=0$ (which is also discussed in Ref. \cite{nabwall}), the  mass degeneracy of the spectrum is
bigger (the particles fit into $\mathcal{N}=2$ hypermultiplets). The parameter $\eta_0$ affects only the masses of the
particles which are in the same $\mathcal{N}=2$ hypermultiplets as the $U(1)$ vector field; $\eta_3$ affects the mass
of the particles which are in the non-Abelian vector hypermultiplet. The case of $\mathcal{N}=2$ SQED was studied in
Ref.\cite{vy}; the results are very similar to our $U(1)$ subsector.

In the limit $\eta_0 e_0\gg\sqrt{\xi}$, we find $M_{S1}^2 \approx 2e_0^4\eta_0^2$ and $M_{S2}^2 \approx \xi^2/(4
\eta_0^2)$. In a similar way, if $\eta_3 e_3 \gg \sqrt{\xi}$ the masses become $M_{T1}^2 \approx 2e_3^4\eta_3^2$ and
$M_{T2}^2 \approx  \xi^2/(4 \eta_3^2)$. The particles with masses $M_{S1}$ and $M_{T1}$ (which  in this limit
correspond to the fields $a$ and $a^k$)
become 
very massive and decouple from the low energy physics.

For $\eta_0 e_0\ll\sqrt{\xi}$ and $\eta_3 e_3\ll\sqrt{\xi}$ nonetheless we find
\beq M_{S1,S2}^2= \xi e_0^2 \pm  \sqrt{\xi} \eta_0 e_0^3, \,\,\,
 M_{T1,T2}^2= \xi e_3^2 \pm  \sqrt{\xi} \eta_3 e_3^3. \eeq
Some of the scalars become slightly heavier and some slightly lighter.

The mass eigenvectors take a quite complicated form for small $\eta_{0,3}$, with at non-trivial mixing between $Q$,
$\tilde{Q}$ and $a,a^k$. On the contrary they are  quite simple for large $\eta_{0,3}$, because the fields $a,a^k$
decouple from the low energy physics. The effective Lagrangian for large $\eta_{0,3}$ is discussed in Appendix A.

\section{$(p,k)$ Coincident Vortices}

\subsection{Second order equations}

In this section, we will study some special solutions representing coincident vortices that live in an Abelian subgroup
of the fields of the theory. These solutions are parameterized
 by two positive integers $(p,k)$; the topological $\mathbb{Z}$ winding number is given by $w=p+k$. This kind of solution gives us the most general vortex with winding $w=1$ up to a
colour-flavour rotation\footnote{As we will discuss in the next section, for higher winding,  we know that it is not
the most general solutions, at least in the BPS case \cite{ht1,hash,EINOS,composite,EKMNOVY}.}.

Due to the symmetry between $Q$ and $\tilde{Q}^\dagger$, for the vortex solution we can consistently set
$\tilde{Q}=Q^\dagger$ \footnote{This can be checked using the variables: $Q_S=(Q+\tilde{Q}^\dagger)/2$,
$Q_D=(Q-\tilde{Q}^\dagger)/2$. With this variables we easily see that $\left. \frac{\partial V}{\partial
Q_D}\right|_{Q_D=0}=0$.}. With this assumption the Euler-Lagrange equations of the theory are:
\begin{eqnarray}
\nonumber  \partial_\mu F^{\mu \nu}_0 &=& e_0^2
\Tr \left( i Q^\dagger (\nabla^\nu Q)-i(\nabla^\nu Q)^\dagger Q  \right), \\
\nonumber  \partial_\mu F^{\mu \nu}_k+ \epsilon_{klm} A_{l \mu} F_{m}^{\mu \nu} &=& e_3^2
\Tr \left( i Q^\dagger \tau_k (\nabla^\nu Q)-i(\nabla^\nu Q)^\dagger \tau_k Q  \right) -\\
\nonumber &-& \epsilon_{klm} ((D^\nu a)^\dagger_l a_m +(D^\nu a)_l a_m^\dagger ), \end{eqnarray}
\beq
 \nabla^\mu \nabla_\mu Q=-\frac{\delta V}{\delta Q^\dagger}, \,\,\,
 \partial^\mu \partial_\mu a=-\frac{\delta V}{\delta a^\dagger},\,\,\,
  D^\mu D_\mu a_k=-\frac{\delta V}{\delta a_k^\dagger}.
  \label{feq} \eeq
We make the following axial symmetric ansatz
(which in the BPS limit reduces to the one of Ref.~\cite{nab}):
 \[ Q=\left(\begin{array}{cc}
\phi_1 e^{p i \varphi} & 0\\
0 & \phi_2  e^{k i \varphi} \\
\end{array}\right),
\] \[  A^3_i=-\frac{\epsilon_{ij} x_j}{r^2} [(p-k)-f_3(r)], \,\,\, A^0_i=-\frac{\epsilon_{ij} x_j}{r^2} [(p+k)-f_0(r)],
\]
\beq a_0=\lambda_0(r), \,\,\, a_3=\lambda_3(r),\,\,\, a_1=a_2=0. \label{ansa}
\eeq
Notice that the adjoint fields $a,a^k$ are
non-trivial in the non-BPS model, whereas they vanish everywhere when $\eta_{0,3}$ are zero (BPS).

The vacuum of the theory is invariant under the following global colour-flavour
locked rotations ($U\in SU(2)_{C+F}$):
\beq Q \rightarrow U Q  U^\dagger,
 \,\,\, \tilde{Q}  \rightarrow U^\dagger Q  U, \,\,\,
 a^k \tau_k \rightarrow U (a^k \tau_k) U^\dagger, \,\,\,
F_{\mu \nu}^k \tau_k \rightarrow U (F_{\mu \nu}^k \tau_k) U^\dagger.\label{colflavsymm}
\eeq
Let us introduce the $S^2$ coordinate $n^k$, with $k=1,2,3$ and $|\vec{n}|=1$:
\beq n^k \tau^k=U \tau^3 U^\dagger.\label{cp1coord}\eeq
Using the parametrization introduced in Eq.~(\ref{cp1coord})
we can write down the expression for a $w=1$ vortex with generic
orientation $n^k$: \[ A^0_i=-\frac{\epsilon_{ij} x_j}{r^2} [1-f_0], \,\,\, A^k_i=-\frac{\epsilon_{ij} x_j}{r^2} [1-f_3]
n^k, \,\,\, a^0=\lambda_0, \,\,\,  a^k= n^k \lambda_3,
\]
\beq Q=\tilde{Q}^\dagger=\frac{\phi_1 e^{I \varphi} +\phi_2}{2} {\mathbf 1}+ \frac{\phi_1 e^{I \varphi} -\phi_2}{2}
\tau^k n^k. \label{eq:single} \eeq It is easy to see that the $(1,0)$ vortex partially break the symmetry in
Eq.~(\ref{colflavsymm}); as a consequence, this object has some internal zero modes associated to this breaking. In
fact, the vortex leaves a  $U(1)$ subgroup of $SU(2)_{C+F}$ unbroken, so that zero modes parameterize a
 $\mathbb{CP}^{1}=SU(2)/U(1)=S^2$.

The energy with respect to $\phi_{1,2},f_{0,3}$ and $\lambda_{0,3}$ is expressed as \[ \mathcal{E} = 2 \pi \int r dr
\left( \frac{f_0'^2}{2  e_0^2 r^2}+\frac{f_3'^2}{2  e_3^2 r^2} +\frac{\lambda_0'^2}{e_0^2}+\frac{\lambda_3'^2}{e_3^2}+
2(\phi_1'^2+\phi_2'^2)+ \right. \]
\[  \left. + \frac{(\phi_1^2+\phi_2^2)(f_0^2+f_3^2)+
2 f_3 f_0 (\phi_1^2-\phi_2^2) }{2 r^2}+ \frac{e_0^2}{2} (\phi_1^2+\phi_2^2-\xi+2 \eta_0 \lambda_0)^2+ \right.\] \beq
\left. + \frac{e_3^2}{2} (\phi_1^2-\phi_2^2+2 \eta_3 \lambda_3)^2 + ((\lambda_0+\lambda_3)\phi_1)^2+
((\lambda_0-\lambda_3)\phi_2)^2 \right). \label{eq:energy}\eeq We have to minimize this expression with the appropriate
boundary conditions for each $(p,k)$:
\begin{eqnarray}
 \nonumber f_3(0) = p-k,\,\,\, f_0(0)=p+k, \,\,\, f_3(\infty)=0, \,\,\,   f_0(\infty) = 0. \\
  \phi_1(\infty) = 1,\,\,\,\,\,\,\,\,\,\, \phi_2(\infty) = 1, \,\,\,\,\,\,\,\,\, \lambda_0(\infty) = 0, \,\,\,   \lambda_3(\infty) = 0.
\end{eqnarray}
We also find for small $r$:
\beq
 \phi_1 \propto {\cal O}(r^p), \,\,\, \phi_2 \propto {\cal O}(r^k), \,\,\, \lambda_0
\propto {\cal O}(1), \,\,\, \lambda_3 \propto {\cal O}(1). \eeq
 The Euler-Lagrange equations obtained are:
\[ \frac{f_0''}{r}-\frac{f_0'}{r^2}=\frac{e_0^2}{r}(f_3 (\phi_1^2-\phi_2^2) +f_0 (\phi_1^2+\phi_2^2) ),\]
\[\frac{f_3''}{r}-\frac{f_3'}{r^2}=
\frac{e_3^2}{r}(f_3 (\phi_1^2+\phi_2^2) +f_0 (\phi_1^2-\phi_2^2) ), \]
\[\phi_1''+\frac{\phi_1'}{r}-\frac{\phi_1(f_0+f_3)^2}{4 r^2}=\]
 \[ =\frac{\phi_1
\left((\lambda_0+\lambda_3)^2+e_0^2 (\phi_1^2+\phi_2^2-\xi+2 \eta_0 \lambda_0)+ e_3^2 (\phi_1^2-\phi_2^2+2 \eta_3
\lambda_3) \right)}{2}, \]
\[\phi_2''+\frac{\phi_2'}{r}-\frac{\phi_2(f_0-f_3)^2}{4 r^2}=\]
\[ =\frac{\phi_2
\left((\lambda_0-\lambda_3)^2+e_0^2 (\phi_1^2+\phi_2^2-\xi+2 \eta_0 \lambda_0)- e_3^2 (\phi_1^2-\phi_2^2+2 \eta_3
\lambda_3) \right)}{2}, \]
\[\lambda_0''+\frac{\lambda_0'}{r}=\frac{
e_0^2\left((a_0+a_3) \phi_1^2 + (a_0-a_3) \phi_2^2 + e_0^4 \eta_0 (\phi_1^2+\phi_2^2-\xi+2 \eta_0 \lambda_0)
  \right)}{2},\]
\beq\lambda_3''+\frac{\lambda_3'}{r}=\frac{ e_3^2\left((a_0+a_3) \phi_1^2 - (a_0-a_3) \phi_2^2 + e_3^4 \eta_3
(\phi_1^2-\phi_2^2+2 \eta_3 \lambda_3)
  \right)}{2}.\label{2ordine}\eeq
 It is easy to check that
these equations can be obtained substituting the ansatz (\ref{ansa})
in Eqs.~(\ref{feq}).
This shows that the ansatz is consistent.

In the following, we will concentrate our effort on the study of the sectors with topological winding $1$ and $2$; in
other words, we will discuss the $(1,0)$, the $(1,1)$ and the $(2,0)$ vortices. In the BPS limit ($\eta_0=\eta_3=0$)
the tension is proportional to the topological winding number ($T_{(1,0)}=2 \pi \xi$, $T_{(1,1)}=T_{(2,0)}=4 \pi \xi$).
For non-BPS solutions, $\eta_{0,3}\neq0$, we find that the tension is always less than  the BPS limit. This is because
the non-BPS terms in the tension formula Eqs.~(\ref{eq:energy}) do not give any contribution if we put the BPS
solutions into the expression ($\lambda_{0,3}$ are identically zero for the BPS solutions). The non-BPS solutions will
 of course be a true minimum or saddle point of the energy functional, so that their energy will be smaller than that of
the BPS configurations\footnote{This also has a clear resemblance to the Abelian case, where type I vortices have a
smaller energy with respect to the BPS case.}.

For fixed $\xi$, the tension of the $(1,1)$ vortex is a function of only $e_0,\eta_0$, because for this vortex $f_3=0$
and $\phi_1=\phi_2$. This is clearly explained by the fact that the $(1,1)$ vortex is completely Abelian. On the other
hand, the tension of the $(1,0)$ and of the $(2,0)$ vortex is a non-trivial function of all the parameters $e_{0,3},
\eta_{0,3}$.

If we take $e_3=e_0$ and $\eta_3=\eta_0$ the vortex becomes easier to study. In this case we can use a more convenient
basis for the gauge field, which is just the sum and the difference of $A_\mu^0$ and $A_\mu^3$. The potential $V$ also
factorize, and takes the form $V=V_1(\phi_1)+V_2(\phi_2)$. Each diagonal component of $Q$ does not interact with the
other ones, and can be treated as an Abelian vortex. For the $(1,0)$ vortex we can use the simple ansatz\footnote{
Notice that we cannot impose $\phi_2=\sqrt{\xi/2}$ even for $(1,0)$ vortex in the generic models.}
\beq Q=\left(\begin{array}{cc}
\phi e^{ i \varphi} & 0\\
0 & \sqrt{\xi/2} \\
\end{array}\right),
\eeq
while for the $(1,1)$ and for the $(2,0)$ vortices we can use
\beq Q=\left(\begin{array}{cc}
\phi(r_1) e^{ i \varphi_1} & 0\\
0 & \phi(r_2) e^{ i \varphi_2} \\
\end{array}\right), \,\,\, Q=\left(\begin{array}{cc}
\phi e^{2 i \varphi} & 0\\
0 & \sqrt{\xi/2} \\
\end{array}\right).
\label{eq:11vor}
\eeq
The system reduces to the Abelian vortex studied in Ref. \cite{vy}. The tension of the $(1,1)$ vortex is exactly twice
the tension of the $(1,0)$ one. In each of the $U(1)$ factors, we have type I superconductivity. Since the two $U(1)$
subgroups are decoupled, the $(1,0)$ and $(0,1)$ vortices do not interact. Furthermore, the tension of the $(2,0)$
vortex is less than twice the tension of the $(1,0)$ vortex.



\subsection{Numerical Solutions}

\begin{figure}[h]
\begin{center}
$\begin{array}{c@{\hspace{.2in}}c@{\hspace{.2in}}c} \epsfxsize=1.5in
\epsffile{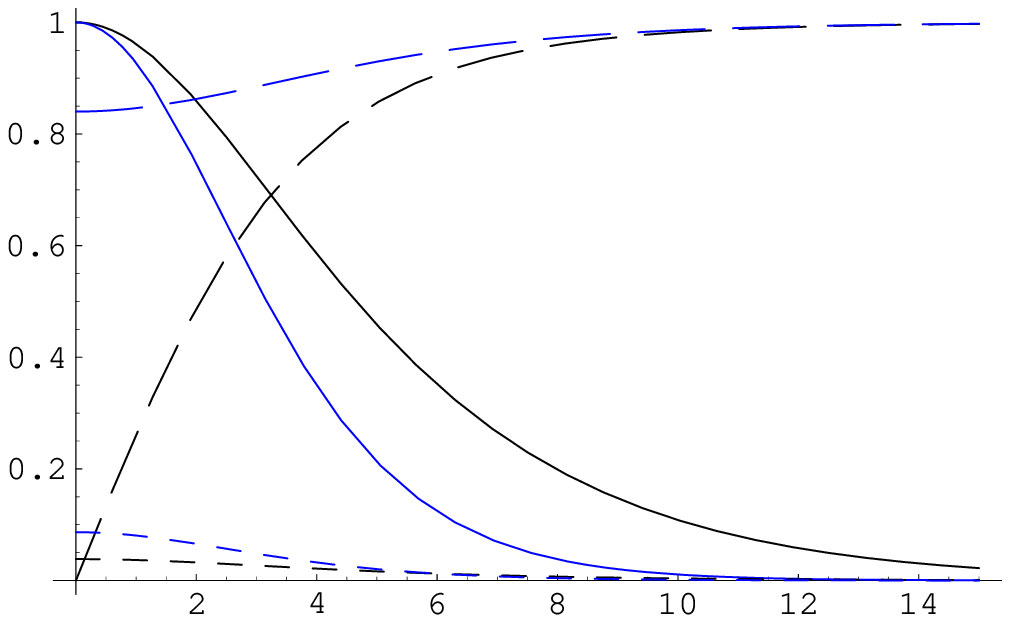} &
    \epsfxsize=1.5in
    \epsffile{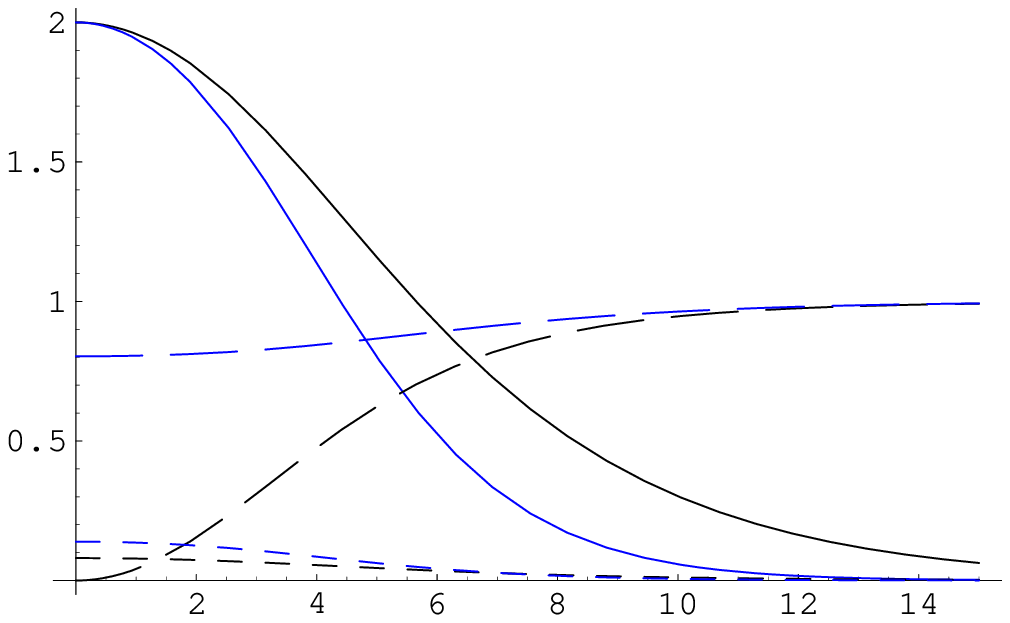} &
    \epsfxsize=1.5in
    \epsffile{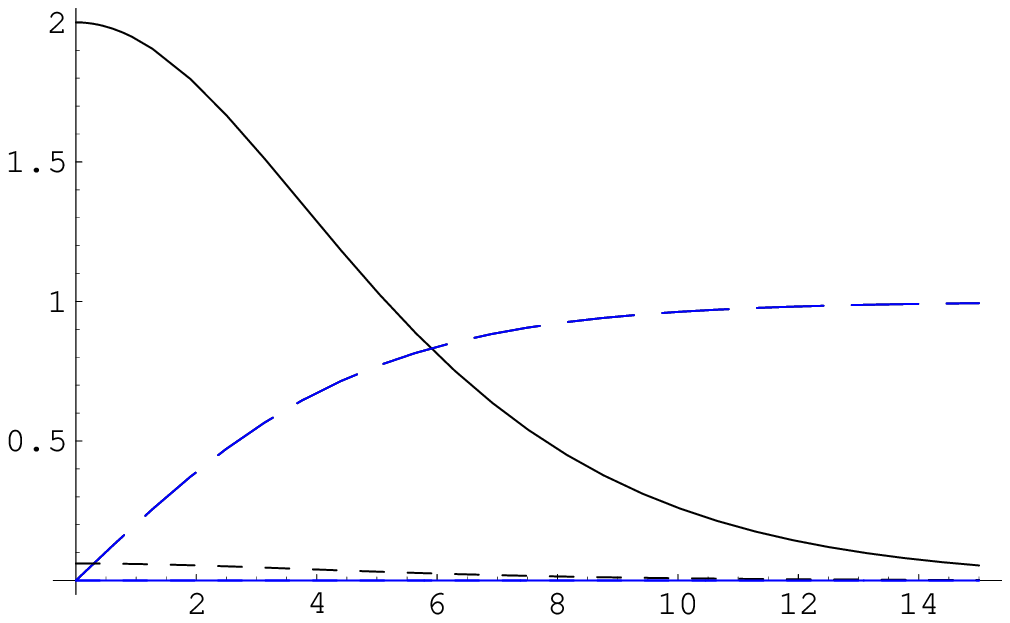}
\end{array}$
\end{center}
\caption{\footnotesize Profile functions $f_0$ (solid black), $f_3$ (solid blue), $\phi_1$ (long dashes, black),
$\phi_2$ (long dashes, blue), $\lambda_0$ (short dashes, black), $\lambda_3$ (short dashed,blue) for the numerical
values $\xi=2,e_0=1/4,e_3=1/2,\eta_0=\eta_3=1$. In the left panel are shown the profiles for the $(1,0)$ vortex, in the
middle, the ones for the $(2,0)$ and in the right, the ones for the $(1,1)$. Note that in this last case
$f_3=\lambda_3=0$ and $\phi_1=\phi_2$.} \label{profili}
\end{figure}

At generic $e_{0,3}, \eta_{0,3}$ Eqs.~(\ref{2ordine}) have been solved numerically. It is a little subtle to solve this
system of ordinary differential equations directly. The difficulties basically arise because there are many equations;
there are also subtleties in defining the boundary conditions at $\infty$, because, in general, the fields which appear
in our  ansatz do not correspond to mass eigenstates. In order to perform the numerics we found that the method of
relaxation is very effective. We add an auxiliary time dependence to the profile functions
$\vec{u}=(f_0,f_3,\phi_1,\phi_2,\lambda_0,\lambda_3)$ . At $t=0$ we start with some arbitrary functions
$u_j(r,0)$\footnote{The choice of the initial conditions is crucial to find convergence.}; the evolution in $t$ is then
given by:
\beq \frac{\partial u_j}{\partial t}= E_{j}. \eeq
If the solution converges with time to a static configuration, then at final time we have obtained a solution of the
equations $E_{j}=0$, which are equivalent to Eqs.~(\ref{2ordine}). The results for  $(p,k)=(1,0),(2,0),(1,1)$ are shown
in Fig.~\ref{profili}.

\begin{figure}[h]
\begin{center}
$\begin{array}{c@{\hspace{.2in}}c} \epsfxsize=2.5in
\epsffile{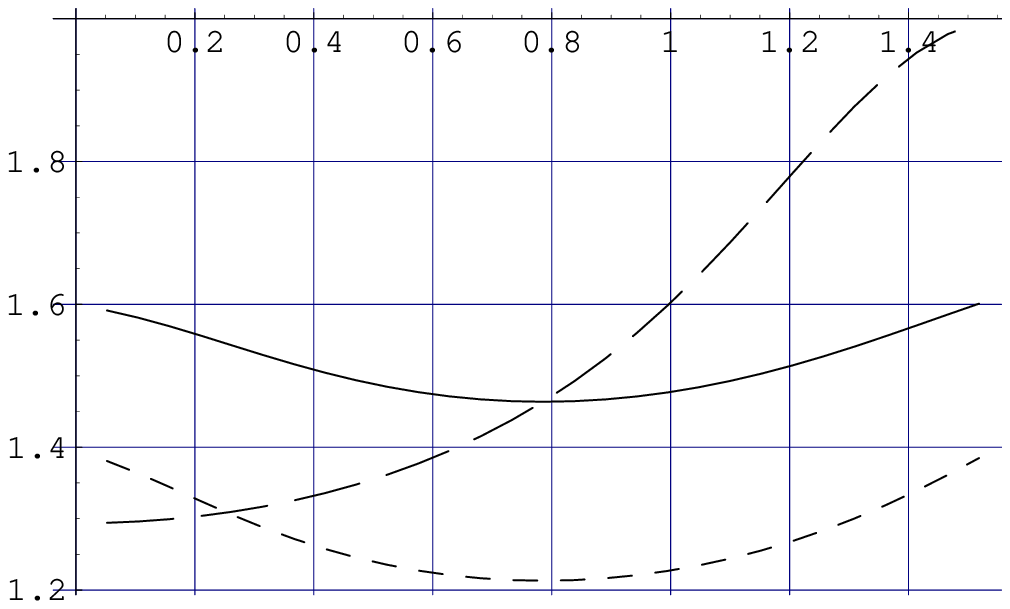} &
    \epsfxsize=2.5in
    \epsffile{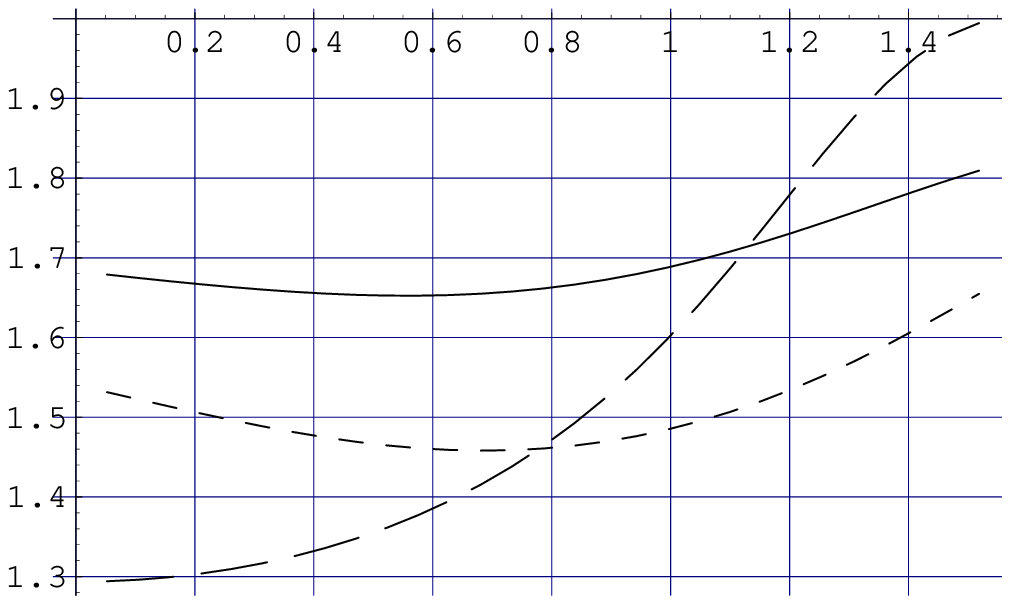}
\end{array}$
\end{center}
\caption{\footnotesize $T_{1,1}$ (long dashes), $T_{2,0}$ (short dashes) and $2 \,T_{1,0}$ (solid)
 for different values of $0<\omega<\pi/2$, where $\eta_3=\eta \sin \omega, \eta_0=\eta \cos \omega$.
 The tension of the BPS 2-vortex is normalized to $T_{BPS}=2$.
 In the left figure the numerical values $e_0=e_3=1/2$,
 $\eta=4$, $\xi=2$ are used; in the right figure
 $e_0=1/2$, $e_3=1/4$, $\eta=4$, $\xi=2$).}
\label{titti}
\end{figure}

It is interesting to compare numerical result for the tension $2T_{1,0}, T_{2,0}$ and $T_{1,1}$ of the $2\times(1,0)$,
$(2,0)$ and $(1,1)$ vortices, respectively (see Fig.~\ref{titti}). We have always found that $T_{2,0}<2 T_{1,0}$. This
is consistent with the fact that at large separation distance, the force between two vortices with the same
colour-flavour orientation is always attractive (we will discuss this aspect in Sect.~4). As can be checked in
Fig.~\ref{titti}, three different regimes have been found for $T_{1,1}$: $T_{1,1}<T_{2,0}<2 T_{1,0}$ or
$T_{2,0}<T_{1,1}<2 T_{1,0}$ or $T_{2,0}<2 T_{1,0}< T_{1,1}$.

If $\eta_0=0, \eta_3 \neq 0$, the tension of the $(1,1)$ vortex is found to be
the same as in the BPS case.
 The tensions
of the $2\times(1,0)$ and of the $(2,0)$ vortex are strictly less than that of the BPS vortices ($\eta_3=0$). Hence in
this case  $T_{2,0}<2 T_{1,0}< T_{1,1}$. On the contrary, if $\eta_3=0, \eta_0 \neq 0$, we have found $T_{1,1}<
T_{2,0}<2 T_{1,0}$ for all the numerical values of the couplings that we have investigated.

\section{Generic Coincident Vortices}

\subsection{The BPS case}

The number of dimensions of the $k$-vortex moduli space in $U(N_c)$ $\mathcal{N}=2$ gauge theory with $N_f=N_c=N$
hypermultiplets has been computed in Ref.~\cite{ht1}. The calculation uses the index theorem and the result is $2 k N$.
Thus for $\eta_0=\eta_3=0$, the moduli space
 of a 2-vortex configuration
is a manifold with eight real dimensions. Two of these dimensions are associated with the global position of the
system; other 2 coordinates are associated with the relative position $R$ of the two elementary vortices. The remaining
4 coordinates are associated with the orientation of the system in the colour-flavour space. In this section, we will
write an ansatz for the case of coincident vortices ($R=0$), and we will show that it is non-trivially consistent with
the second order equations of the theory. We will correct a technical mistake in Ref.~\cite{composite}, where the
problem was studied using first order BPS equations.

Modulo a global $SU(2)$ rotation we can parameterize a subset of the moduli space with the angle $\alpha$ between
$\vec{n}_1$ and  $\vec{n}_2$. The expression for $Q=\tilde{Q}^\dagger$ is\footnote{The ansatz for $Q$ used in
Ref.~\cite{composite} is: $\kappa_1=\kappa \, z_1 \, z_2$, $\kappa_2=\kappa \, z_1$, $\kappa_3=\kappa \, z_2$,
$\kappa_4=\kappa$. This form is not sufficiently general, because we have to keep all the four squark components
independent variables; we can show numerically that $\kappa_1 \kappa_4 \neq \kappa_2 \kappa_3$. Moreover, the profile
function $h(r)$ introduced in the same paper turns out to be zero: as a consequence, the correct ansatz takes a much
simpler form. The conclusions of Ref.~\cite{composite} about the vortex moduli space although are not changed by these
technical points.}:
 \beq
 Q =  \left(\begin{array}{cc}
-\cos \frac{\alpha}{2}  e^{ 2 i \varphi}  \kappa_1
& \sin \frac{\alpha}{2}  e^{  i \varphi} \kappa_2  \\[3mm]
- \sin \frac{\alpha}{2}  e^{  i \varphi}  \kappa_3
& -\cos \frac{\alpha}{2} \kappa_4  \\
\end{array}\right). \label{ans1}
\eeq
The ansatz for the gauge fields is: \[ A^0_{(i)} = -\frac{\epsilon_{ij} x_j}{r^2}  (2-f_0), \,\,\,
 A^3_{(i)} = -\frac{\epsilon_{ij} x_j}{r^2}  ((1+\cos \alpha )-f_3),
\]
\beq
A^1_{(i)} = -\frac{\epsilon_{ij} x_j}{r^2} (\sin \alpha) (\cos\varphi)  (1-g), \,\,\,
A^2_{(i)} = +\frac{\epsilon_{ij} x_j}{r^2} (\sin \alpha) (\sin \varphi)  (1-g).
 \label{ans2}\eeq
We have introduced here the profile functions $\kappa_1(r),\kappa_2(r),\kappa_3(r), \kappa_4(r)$
 for the squark scalars
and $f_0(r),f_3(r),g(r)$ for the gauge field. For $r\rightarrow\infty$ all the gauge profile functions vanish and all
the squark ones go to the value $\sqrt{\xi/2}$. The boundary conditions at $r \rightarrow 0$ are:
\beq
f(r)=2+\mathcal{O}(r^2), \,\,\,
 f_3(r)=(1+\cos \alpha) + \mathcal{O}(r^2), \,\,\,
 g(r)=1+\mathcal{O}(r^3),
 \eeq
\[\kappa_1(r)\to  \mathcal{O}(r^2),\,\,\, \kappa_2(r)\rightarrow \mathcal{O}(r), \,\,\,
\kappa_3(r)\rightarrow \mathcal{O}(r), \,\,\, \kappa_4(r)\rightarrow
\mathcal{O}(1)\,. \]
For the $(2,0)$ vortex we have:
\[ \alpha=0, \,\,\, \phi_1= \kappa_1, \,\,\,
 \phi_2=\kappa_4,\]
while for the $(1,1)$ vortex (after a simple diagonalization):
\[ \alpha=\pi, \,\,\, \phi_1= \kappa_2= \phi_2=\kappa_3.\]

\begin{figure}[h]
\begin{center}
$\begin{array}{c@{\hspace{.2in}}c} \epsfxsize=2.5in
\epsffile{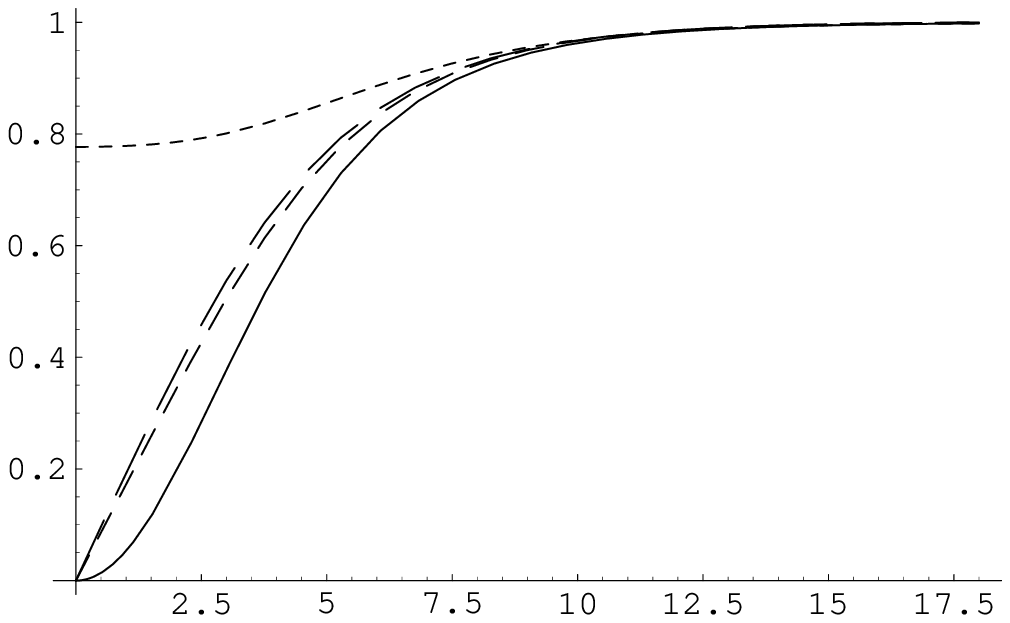} &
    \epsfxsize=2.5in
    \epsffile{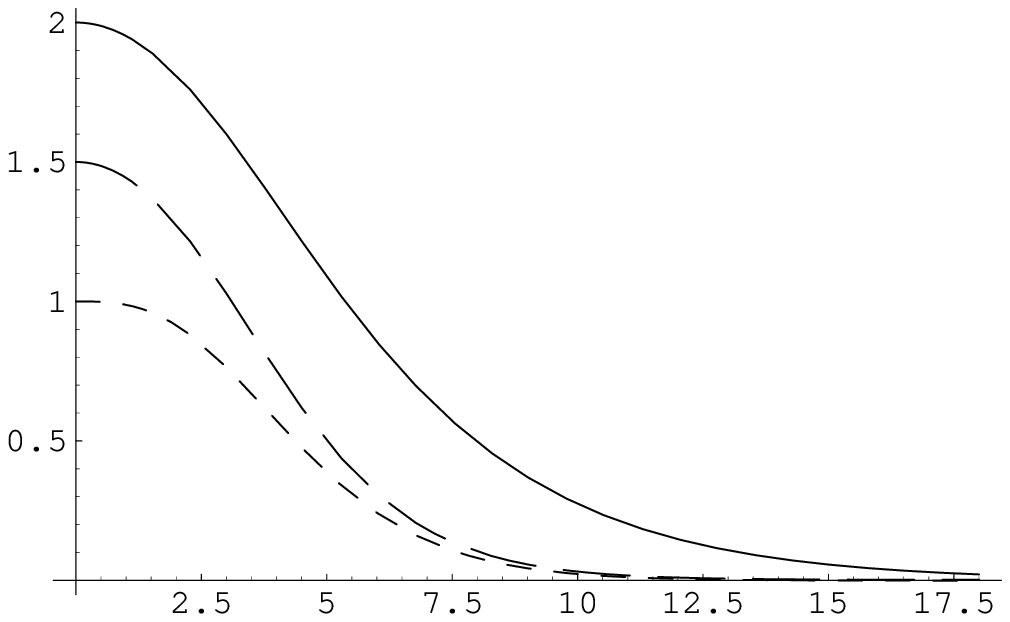}
\end{array}$
\end{center}
\caption{\footnotesize Vortex profile functions for $\alpha=\pi/3$. In the left panel there are $\kappa_1$ (solid),
$\kappa_2$ (long dashes), $\kappa_3$ (short dashes), $\kappa_4$ (dots);
 in the right panel there are $f_0$ (solid), $f_3$ (long dashes), $g$ (short dashes).
The following numerical values have been used: $\xi=2$,
 $e_0=1/4$, $e_3=1/2$).}
\label{alphabps}
\end{figure}

For the BPS vortex it is simpler to consider first order equations; but we are interested in understanding what is
happening for $\eta_{0,3}\neq 0$. Thus we will write the equations in a form that can be easily generalized to a
non-BPS setting. This will also give the possibility to check our equations and numerical results, just comparing the
result for the tensions against the exact Bogomol'nyi bound; for completeness, we provide the  first order BPS
equations in Appendix~B. The energy density due to the kinetic part of the gauge field is:
  \beq S_g=
\frac{{f_0'}^2}{2\,r^2\,{{e_0}}^2} + \frac{f_3'^2}{2\,r^2\,{{e_3}}^2}+
 \frac{{\sin^2 \alpha }\,{g'}^2}{2\,r^2\,{{e_3}}^2}. \eeq
The part due to the kinetic energy of squark is: \[ S_Q=  \cos^2 \frac{\alpha }{2}\,(\kappa_1'^2+\kappa_4'^2 )+
 \sin^2 \frac{\alpha }{2}\,(\kappa_2'^2+\kappa_3'^2 ) + \]
\[+\cos^2 \frac{\alpha }{2}\,
\left( \frac{(1-\cos \alpha +f_0+f_3)^2 \kappa_1^2}{4 r^2}+
\frac{(1-\cos \alpha -f_0+f_3)^2 \kappa_4^2}{4 r^2}   \right)+ \]
\[+\sin^2 \frac{\alpha }{2}\,
\left( \frac{(1+\cos \alpha -f_0-f_3)^2 \kappa_2^2}{4 r^2}+
\frac{(1+\cos \alpha +f_0-f_3)^2 \kappa_3^2}{4 r^2}   \right)+ \]
\[ + \frac{(1-g)^2 \sin^2 \alpha}{4 r^4}
\left( \cos^2 \frac{\alpha }{2}\, (\kappa_1^2+\kappa_4^2)+
\sin^2 \frac{\alpha }{2}\, (\kappa_2^2+\kappa_3^2)
 \right)-
\]
\beq  -\frac{(1-g) \sin^2 \alpha}{2 r^4} \left( (1+f_0) \kappa_1 \kappa_3 + (1-f_0) \kappa_2 \kappa_4 \right).
\eeq
The part due to the potential reads:
 \[ V_{BPS}=\frac{e_0^2}{2}
 {\left( \cos^2 \frac{\alpha }{2}\,
 ({{{\kappa }_1}}^2+\,{{{\kappa }_4}}^2) +
  \sin^2 \frac{\alpha }{2}\,
  ({{{\kappa }_2}}^2 + {{{\kappa }_3}}^2)   -\xi  \right) }^2 + \]
 \[+  \frac{e_3^2}{2} \left\{ {\left(\cos^2 \frac{\alpha }{2}
 ({{{\kappa }_1}}^2 -{{{\kappa }_4}}^2 )+
  \sin^2 \frac{\alpha }{2}
 ({{{\kappa }_2}}^2 - {{{\kappa }_3}}^2)  \right) }^2  +
  \sin^2 \alpha \,{\left( {{\kappa }_1}\,{{\kappa }_3} -
  {{\kappa }_2}\,{{\kappa }_4} \right) }^2 \right\}.        \]
  \beq \eeq
The total energy is given by:
\beq \mathcal{E} = 2 \pi \int r dr (S_g+2 S_Q+V_{BPS}).\eeq
It is straightforward to write the Euler-Lagrange equations for this energy density, which are a system of seven second
order equations, one for each profile function; for brevity we will not show them explicitly in the paper. We have
 solved this system numerically with the same method used in Sect.~3.2; in Fig.~\ref{alphabps} is shown an example of the
solution. The tension is found to be equal to $T_{BPS}=4 \pi \xi$ with an excellent precision for every $\alpha$; this
is a good numerical check for the solution obtained.

An analytical check of the ansatz can also be found substituting Eqs.~(\ref{ans1},\ref{ans2})
 into the  Euler-Lagrange equations (\ref{feq}).
With this approach we find a system of the same seven second order equations with the following first order expression:
\beq K=e_3^2 r^2 (\kappa_3 \kappa_1'-\kappa_1 \kappa_3'+
\kappa_2 \kappa_4' - \kappa_4 \kappa_2')-
(1-g) f_3'-(f_3-\cos \alpha) g'=0.
\eeq
This seems to be a paradox, because this is a system of eight
differential equation with seven unknown functions.
Actually, everything is consistent, because using the seven second order
equations we can show the following property:
\beq \frac{d K}{d r}= \frac{K}{r}, \eeq
which shows that $K$ is linear in $r$. From the boundary conditions of the profile functions, we find that the
coefficient of this linear function has to be zero. This shows that our ansatz is consistent with the equations of
motion.

\subsection{The non-BPS case: $\eta_0,\eta_3 \neq 0$}

For $\eta_0,\eta_3 \neq 0$ the $(1,1)$ and the $(2,0)$ vortices are still solutions to the equations of motion; so
these field configurations are extremal points of the energy which can be local minima or saddle points. For generic
values of the parameters, we find $T_{(1,1)}\neq T_{(2,0)}$, so the continuous moduli space interpolating between these
two particular solutions disappears. For small values of $\eta_0,\eta_3$ we expect that the low energy physics of these
solitons is described by an effective potential of the moduli space. In this section, we will estimate this potential
numerically for generic values of $\alpha$.

   A constraint on this potential comes from the BPS limit at $\eta_0,\eta_3=0$.
In this case, a continuous family of degenerate solutions exists, with tension $T=4 \pi \xi$. If we insert these
solutions into the energy density for $\eta_0,\eta_3 \neq 0$, the energy of these field configurations does not change.
However, the solutions to the second order equations have energies which are less than this value. This sets an upper
bound:
\beq T(\alpha)\leq T_{BPS}=4 \pi \xi. \eeq
 There is an obvious invariance of the equations:
\[ \alpha \rightarrow - \alpha. \]
Indeed, if we expand around $\alpha=0+\delta$ or $\alpha=\pi+\delta$ we find that the linear order in $\delta$ is zero
and that the first non-trivial correction to the tension is $\mathcal{O}(\delta^2)$. This shows that solutions with
$\alpha=0,\pi$ correspond to local minima or maxima of the tension. In order to find which of the two alternatives holds,
an explicit calculation is needed.

In order to compute the potential of the vortex moduli space,
 we generalize the ansatz
that we have used for the solutions in the BPS case, using the same expressions for the gauge fields, $Q$ and the
following expression for the adjoint fields: \[ a_0=\lambda_0(r),\,\,\, a_3=\lambda_3(r), \] \beq a_1= (\sin \alpha)
\frac{x_1}{r} \lambda_{12}(r), \,\,\, a_2= (\sin \alpha) \frac{-x_2}{r} \lambda_{12}(r), \eeq where we have introduced
the profile functions $\lambda_0,\lambda_3,\lambda_{12}$, with the following boundary conditions:
\beq
 \lambda_0(\infty) = 0,\,\,\, \lambda_3(\infty) = 0, \,\,\, \lambda_{12}(\infty) = 0,
 \eeq
and the following $r\rightarrow 0$ behaviour:
\beq
 \lambda_0 \propto {\cal O}(1), \,\,\,  \lambda_3 \propto {\cal O}(1), \,\,\, \lambda_{12} \propto {\cal O}(r). \eeq
 This ansatz is suggested by the expression we get for these adjoint fields in the limit of large $\eta_0,\eta_3$, where
 we can integrate these fields out (see Appendix~A).
In the following we replace these expressions in the action and find second order equations for the profile functions
for generic $\alpha$. These field configurations at $\alpha\neq0,\pi$ are not solutions to the full equations of
motion, Eqs.~(\ref{feq}); they are just functional generalizations of the $BPS$ solutions. We use these profiles as
reasonable test functions to compute the effective moduli space potential.
\begin{figure}[h]
\begin{center}
$\begin{array}{c@{\hspace{.2in}}c@{\hspace{.2in}}c} \epsfxsize=1.6in
\epsffile{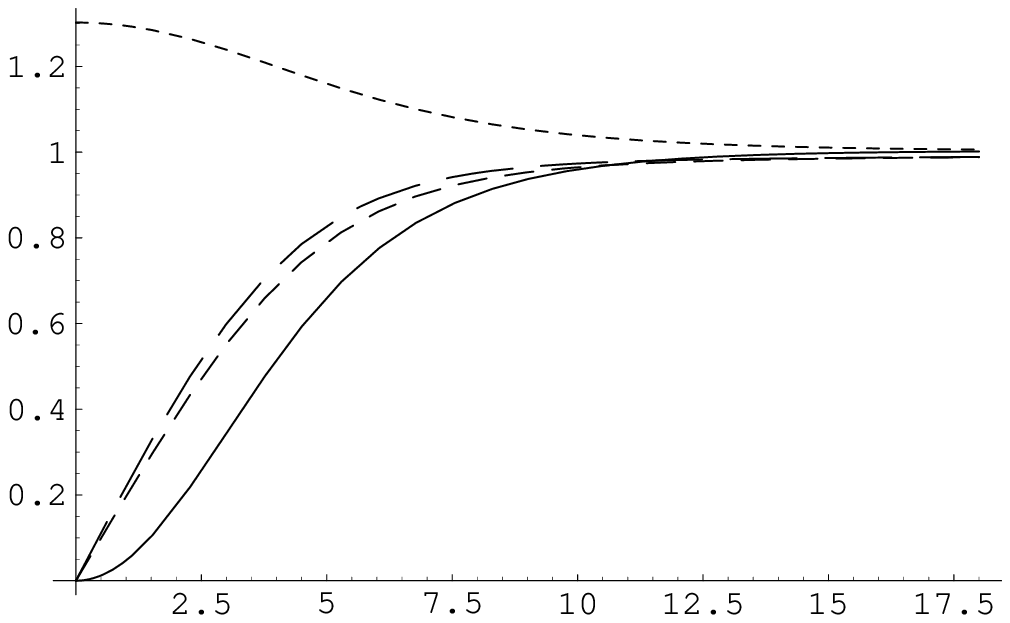} &
    \epsfxsize=1.6in
    \epsffile{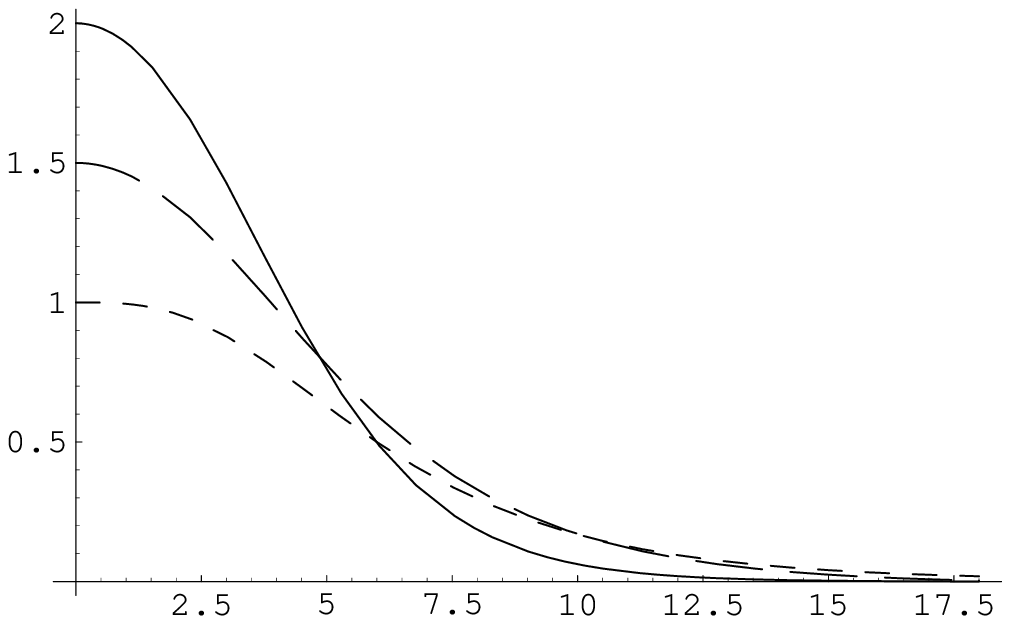} &
    \epsfxsize=1.6in
    \epsffile{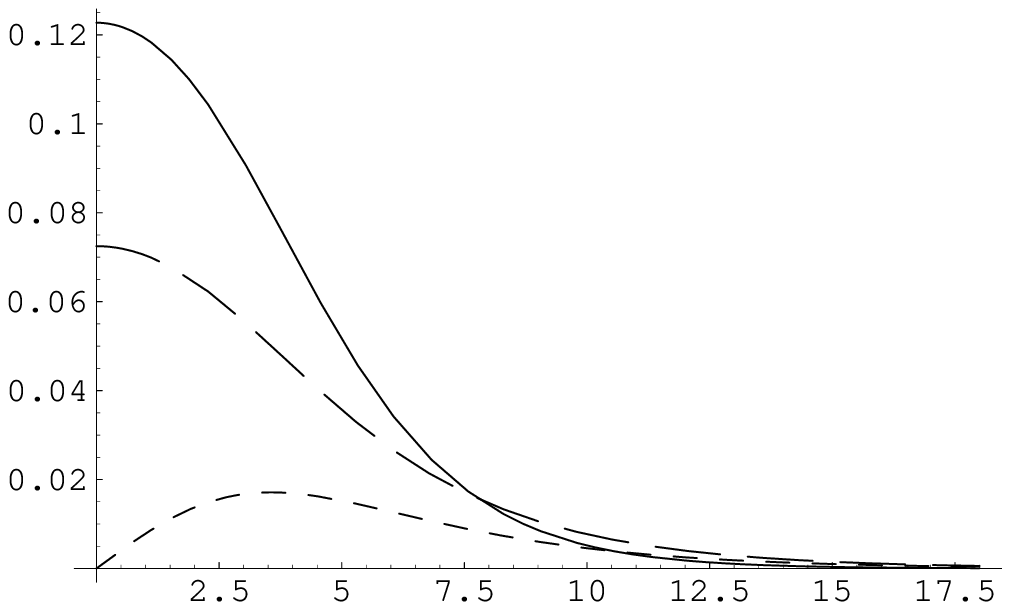}
\end{array}$
\end{center}
\caption{\footnotesize Vortex profiles for $\alpha=\pi/3$ and $\xi=2$, $e_0=1/2$, $e_3=1/4$, $\eta_0=\eta_3=1$. In the
left panel there are $\kappa_1$ (solid), $\kappa_2$ (long dashes), $\kappa_3$ (short dashes), $\kappa_4$ (dots);
 in the middle panel $f_0$ (solid), $f_3$ (long dashes), $g$ (short dashes); in the right panel
$\lambda_0$ (solid), $\lambda_3$ (long dashes), $\lambda_{12}$ (short dashes).}
\label{pralpha}
\end{figure}

The kinetic energy of the adjoint scalars is: { \[ S_a= \frac{{{{\lambda }_0}'}^2}{{{e_0}}^2} +
 \frac{{{{\lambda }_3}'}^2 + {\sin (\alpha )}^2\,{{{\lambda }_{12}}'}^2 }{e_3^2}
+ \frac{{\sin (\alpha )}^2}{r^2 e_3^2} \left\{
   (1 -g)^2  \,      {{{\lambda }_3}}^2 + \right. \]
\beq
\left. +
        ( f_3 - \cos (\alpha ) ) ^2  \,      {{{\lambda }_{12}}}^2
    +      2\,{{\lambda }_3}\,{{\lambda }_{12}} \,
     \left( \cos \alpha  - f_3 \right) \,\left(  g-1 \right) \,
           \right\}.
           \eeq}

\begin{figure}[h]
\begin{center}
$\begin{array}{c@{\hspace{.2in}}c@{\hspace{.2in}}c} \epsfxsize=1.6in \epsffile{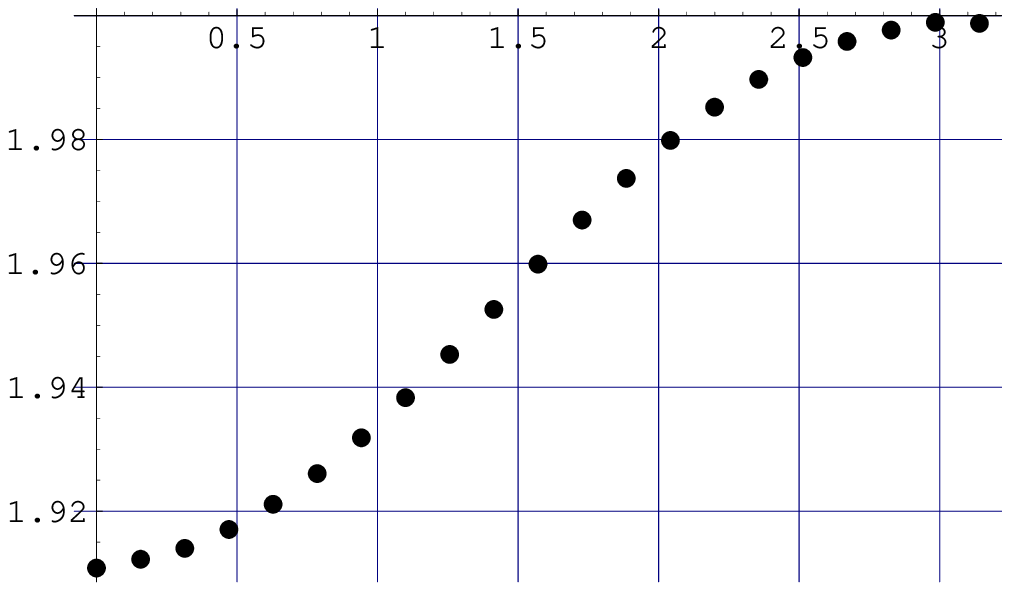} &
    \epsfxsize=1.6in
    \epsffile{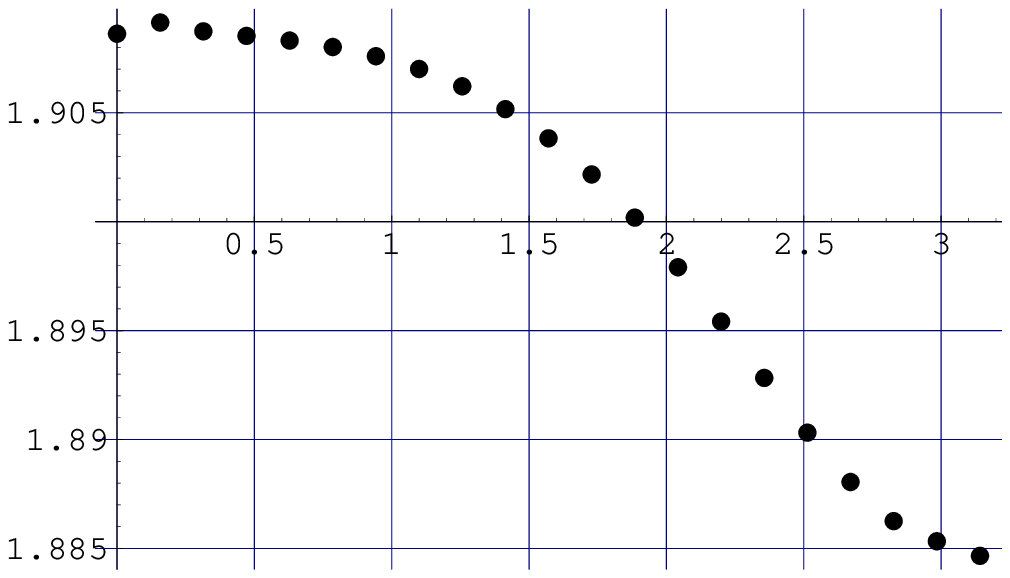} &
    \epsfxsize=1.6in
    \epsffile{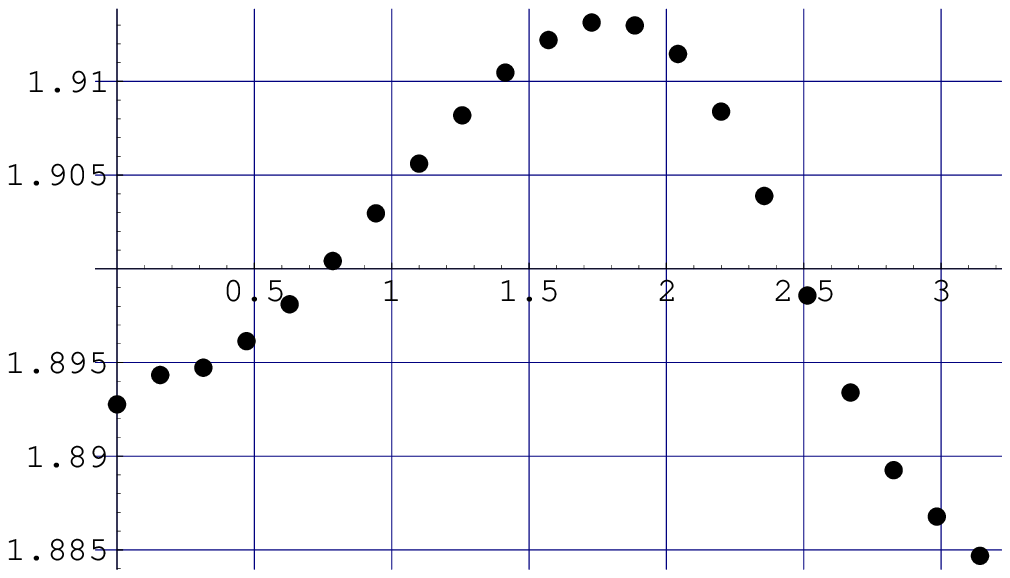}
\end{array}$
\end{center}
\caption{\footnotesize Vortex tension as function of $\alpha$ in the topological winding 2 sector;
 the tension of the BPS 2-vortex is normalized to $T_{BPS}=2$.
  In the left panel: $\xi=2$, $e_0=1/2$, $e_3=1/2$, $\eta_0=0.1$, $\eta_3=1$.
In the middle panel: $\xi=2$, $e_0=1/2$, $e_3=1/2$, $\eta_0=1$, $\eta_3=0.1$. In the right panel: $\xi=2$, $e_0=1/2$,
$e_3=1/4$, $\eta_0=\eta_3=1$.} \label{confronto}
\end{figure}
The potential term is: { \[ V=\frac{e_0^2}{2}
 {\left(  {\cos (\frac{\alpha }{2})}^2\,({{{\kappa }_1}}^2+\,{{{\kappa }_4}}^2) + {\sin (\frac{\alpha }{2})}^2\,({{{\kappa }_2}}^2 + {{{\kappa }_3}}^2)  +
         2\,{{\eta }_0}\,{{\lambda }_0} -\xi  \right) }^2 + \]
 \[+  \frac{e_3^2}{2}\,{\left( {\cos (\frac{\alpha }{2})}^2\,({{{\kappa }_1}}^2 -{{{\kappa }_4}}^2 )+
 {\sin (\frac{\alpha }{2})}^2\,({{{\kappa }_2}}^2 - {{{\kappa }_3}}^2) +
           2\,{{\eta }_3}\,{{\lambda }_3} \right) }^2 +  \]
 \[  +\frac{e_3^2}{2} {\sin (\alpha )}^2\,{\left( {{\kappa }_1}\,{{\kappa }_3} - {{\kappa }_2}\,{{\kappa }_4} +
            2\,{{\eta }_3}\,{{\lambda }_{12}} \right) }^2
  +2\,{\sin (\alpha )}^2\, {{\lambda }_0}\,
     {{\lambda }_{12}}\,(
  {{\kappa }_1}\,{{\kappa }_3}\, - {{\kappa }_2}\,{{\kappa }_4}) + \]
\[+   \left( {\sin (\frac{\alpha }{2})}^2 \,{{{\kappa }_3}}^2 + {\cos (\frac{\alpha }{2})}^2 \,{{{\kappa }_4}}^2 \right) \,
     \left( ({{{\lambda }_0}} -  {{{\lambda }_3}})^2 +
       {\sin (\alpha )}^2\,{{{\lambda }_{12}}}^2 \right)  + \]
\beq+    \left(\,{\cos (\frac{\alpha }{2})}^2\,{{{\kappa }_1}}^2\,+    \,{\sin (\frac{\alpha }{2})}^2\,{{{\kappa
}_2}}^2\, \right)
     \left( ({{{\lambda }_0}} +  {{{\lambda }_3}})^2 +
       {\sin (\alpha )}^2\,{{{\lambda }_{12}}}^2 \right). \eeq
 }
The energy is the sum of all pieces:
\beq \mathcal{E} = 2 \pi \int r dr (S_g+2 S_Q+S_a+V).\eeq

The system of ten second order differential equations which we obtain is quite complicated, but can still be solved
numerically. The qualitative plot of the profile functions is similar to the BPS case (see Fig.~\ref{alphabps} and
Fig.~\ref{pralpha}). The tension is a non-trivial function on the coordinate $\alpha$, which gives us an effective
potential of the moduli space. We solved this equations numerically for different values of the couplings
$e_0,e_3,\eta_0,\eta_3$ and we have found three different regimes (see  Fig.~\ref{confronto}). The tension can have a
maximum at $\alpha=0$ and a minimum at $\alpha=\pi$; we also find the opposite situation in which there is a minimum at
$\alpha=0$ and a maximum at $\alpha=\pi$. The third alternative is that both $\alpha=0,\pi$ are local minima of the
tension, with one of them a metastable minimum. We never obtain a minimum at $\alpha \neq 0, \pi$.

\section{ Vortex Interactions at Large Distance}

\subsection{Vortex profiles at large distance}

At large distance $r$ from the center of the vortex, the equations can be linearized and solved analytically. Let us
introduce the following notation:
\beq \phi_1=\sqrt{\xi/2}+\delta \phi_1, \,\,\,
 \phi_2=\sqrt{\xi/2}+\delta \phi_2, \,\,\,
 \vec{v}=(\delta \phi_1,\delta \phi_2,\lambda_0,\lambda_3). \eeq
The following linear differential equations can be written:
\beq \vec{v}''(r)+\frac{\vec{v}'(r)}{r}-W \vec{v}(r)=0,\eeq
where the matrix $W$ is given by:
\beq W= \left(\begin{array}{cccc}
\frac{\xi (e_0^2+e_3^2)}{2} & \frac{\xi (e_0^2-e_3^2)}{2}&
 \sqrt{\frac{\xi}{2}} e_0^2 \eta_0 &  \sqrt{\frac{\xi}{2}} e_3^2 \eta_3  \\[3mm]
\frac{\xi (e_0^2-e_3^2)}{2} & \frac{\xi (e_0^2+e_3^2)}{2}&
 \sqrt{\frac{\xi}{2}} e_0^2 \eta_0 &  -\sqrt{\frac{\xi}{2}} e_3^2 \eta_3   \\[3mm]
\sqrt{2 \xi} e_0^4 \eta_0 & \sqrt{2 \xi} e_0^4 \eta_0&
 2 \eta_0^2 e_0^4+\xi e_0^2 &  0 \\[3mm]
\sqrt{2 \xi} e_3^4 \eta_3 & -\sqrt{2 \xi} e_3^4 \eta_3&
 0 &   2 \eta_3^2 e_3^4+\xi e_3^2  \\
\end{array}\right).\eeq
The eigenvalues of the matrix $W$ are in direct correspondence with some
of the scalar spectrum of the theory (see Eq.~(\ref{masse})):
\beq w_{1,2}=M^2_{S1,S2}= \xi e_0^2 + e_0^4 \eta_0^2\pm  \sqrt{2 \xi \eta_0^2 e_0^6+e_0^8 \eta_0^4}, \eeq
\[ w_{3,4} =M^2_{T1,T2}= \xi e_3^2 +  e_3^4 \eta_3^2 \pm  \sqrt{2 \xi \eta_3^2 e_3^6+e_3^8 \eta_3^4}.\]
The corresponding eigenvectors are:
{\small \[ \vec{v}_{1,2}=\left( \frac{- {{e_0}}^4 {{{\eta }_0}}^2
\pm {\sqrt{2 \xi {{e_0}}^6 {{{\eta }_0}}^2 + {{e_0}}^8 {{{\eta }_0}}^4}}}
   {2 {\sqrt{2 \xi }}{{e_0}}^4 {{\eta }_0}},
  \frac{- {{e_0}}^4 {{{\eta }_0}}^2   \pm {\sqrt{2 \xi
  {{e_0}}^6 {{{\eta }_0}}^2 + {{e_0}}^8 {{{\eta }_0}}^4}}}
   {2\,{\sqrt{2 \,\xi }}\,{{e_0}}^4\,{{\eta }_0}},1,0\right),
\]
\[ \vec{v}_{3,4}=
\left( \frac{- {{e_3}}^4 {{{\eta }_3}}^2   \pm {\sqrt{2 \xi
{{e_3}}^6 {{{\eta }_3}}^2 + {{e_3}}^8 {{{\eta }_3}}^4}}}
   {2 {\sqrt{2  \xi }} {{e_3}}^4 {{\eta }_3}},
  \frac{{{e_3}}^4 {{{\eta }_3}}^2 \mp {\sqrt{2 \xi {{e_3}}^6
  {{{\eta }_3}}^2 + {{e_3}}^8\,{{{\eta }_3}}^4}}}
   {2 \, {\sqrt{2 \, \xi }}\,{{e_3}}^4\,{{\eta }_3}},0,1\right).
\]}
Note that $\vec{v}_k$ are also
 eigenvectors of the mass matrix defined in Eq.~(\ref{mmassa}).

The solutions to these equations which are zero at infinity are given by
the modified Bessel function:
\beq \vec{v}(r)=  \sum _{k=1,\cdots,4} b_k \vec{v}_k K_0(\sqrt{w_k} r), \eeq
where $b_k$ are appropriate constants which can be found solving the complete differential equation also at small $r$.
For large $x$ we can use:
\beq K_0(x) \approx \sqrt{\frac{\pi}{2 x}} e^{-x}.\eeq
The asymptotic solutions for the scalar profiles read:
\beq \vec{v}(r) \approx \sum _{k=1,\cdots,4} b_k \vec{v}_k \sqrt{\frac{\pi}{2 \sqrt{w_k}r}}
 e^{-\sqrt{w_k} r}. \eeq

The large $r$ equations for $f_3$ and $f_0$ are:
\beq f_0''-\frac{f_0'}{r}-\xi e_0^2 f_0=0, \,\,\,
f_3''-\frac{f_3'}{r}-\xi e_3^2f_3=0. \eeq This leads to the following asymptotic expression in terms of Bessel
functions:
\beq f_{0,3} = c_{0,3} r K_1(e _{0,3} \sqrt{\xi} r)\propto  \sqrt{r} e^{- ( e _{0,3} \sqrt{\xi})  r },\eeq
where $c_0,c_3$ are constants which should be determined by
the original 2nd order differential equations. Note that there is the identity:
$K_1(r)=-K_0'(r)$.

\subsection{Static vortex potential}

The next step is to reproduce the vortex asymptotic interactions in the effective linear theory by coupling the
low-energy degrees of freedom to an effective scalar density $\rho$ and an effective vector current $j_\mu$. We shall
extend an approach used in Refs. \cite{speight,my}.

First of all, we have to discuss the bosonic particle spectrum of the theory. There is a massive $U(1)$ vector and a
massive $SU(2)$ vector; then in principle there are $12$ complex scalar fields ($Q$, $\tilde{Q}$, $A$, $A^k$). For the
vortex solution we have used the ansatz $Q=\tilde{Q}^\dagger$, so in order to discuss the vortex interactions we can
neglect the modes that break this condition\footnote{If we wish to include these extra modes in the low energy theory,
we need only to promote the real fields $S_1,S_2,T_1^k,T_2^k$ in Eq.~(\ref{gentle}) to complex fields.}. There are $16$
real fields, $4$ of which are eaten by the Higgs mechanism; finally we have $12$ physical scalars. We have already
calculated the masses of these particles in Sect.~2.2; at low energy and at low coupling we can write a free theory
which describes the infrared physics:
\beq \mathcal{L}_{free}=\frac{1}{4 e_0^2} (F_{\mu \nu}^0)^2+\frac{\xi}{2}A^0_\mu A^{0 \mu}+
\frac{1}{4 e_3^2} (F_{\mu \nu}^k)^2+\frac{\xi}{2}A^k_\mu A^{k \mu} + \label{gentle} \eeq
\[+ \frac{1}{2} (\partial_\mu S_l)^2+ \frac{1}{2} (\partial_\mu T_l^k)^2+
 \frac{M_{Sl}^2}{2} S_l^2+ \frac{M_{Tl}^2}{2} (T_l^k)^2.
\]
This effective Lagrangian contains three real scalar fields, $S_{l=0,1,2}$, which are $SU(2)$ singlets and three real
scalar fields, $T^k_{l=0,1,2}$, which are $SU(2)$ triplets. These scalars correspond to the appropriate eigenvectors of
the mass matrix in Eq.~(\ref{mmassa}). The index $k$ is an $SU(2)$ triplet index; this $SU(2)$ group corresponds to the
$SU(2)_{C+F}$ in the full theory.

In order to include external sources (=vortices) in this effective Lagrangian, we need the following effective terms:
\beq \mathcal{L}_{source}= \rho_{Sl} S_l +\rho_{Tl}^k T^k_l+j^{0 \mu} A^{0\mu}+
j^{k \mu} A^{k \mu}. \eeq
The corresponding wave equations are:
\beq (\square +M_{Sl}^2) S_l=\rho_{Sl}, \,\,\, (\square +M_{Tl}^2) T_l^k=\rho_{Tl}^k,
\eeq
\[ (\square+{e_0}^2 \xi)A^{0\mu}=j^\mu, \,\,\, (\square+{e_3}^2 \xi) A^{k\mu}=j^{k \mu}. \]
On the other hand, for the $(1,0)$ vortex with orientation $n^k$, we have the following asymptotic profiles, converted
into the singular real $Q$ gauge:
\beq S_0=0,\,\,\, S_1=b_1 K_0(M_{S1} r) \,\,\, S_2=b_2 K_0(M_{S2} r),\eeq
\[ T_0^k=0,\,\,\, T_1^k=b_3 n^k K_0(M_{T1} r) \,\,\, T_2^k=b_4 n^k K_0(M_{T2} r),\]
\[ \vec{A}^0=-c_0(\hat{z}  \wedge \nabla K_0(e_0 \sqrt{\xi} r)), \,\,\,
\vec{A}^k=-c_3 n^k (\hat{z}  \wedge \nabla K_0(e_3 \sqrt{\xi} r)),  \] where $\nabla$ is only the ordinary gradient and
not the covariant derivative as in the other sections. We need the following mathematical identity for the $2+1$
dimensional Laplacian of $K_0$ in term of Dirac's $\delta$ function:
\beq (-\Delta +M^2) K_0(M r)= 2 \pi \delta( \vec{r} ).\eeq
The following expressions are found for the scalar densities corresponding to a vortex placed at the position $\vec{x}$
and having orientation $n^k$:
\beq \rho_{S0}=0, \,\,\,\rho_{S1}=2 \pi b_1 \delta(\vec{x}),
\,\,\,\rho_{S2}=2 \pi b_2 \delta(\vec{x}), \eeq
\[ \rho_{T0}=0, \,\,\,\rho_{T1}=2 \pi b_3 n^k \delta(\vec{x}),
 \,\,\,\rho_{T2}=2 \pi b_4 n^k \delta(\vec{x}). \]
In a similar way we obtain the following expressions for the currents:
\beq \vec{j}=-2 \pi c_0 \hat{z} \wedge \nabla \delta(\vec{x}), \,\,\,
\vec{j}^k=-2 \pi c_3 n^k \hat{z} \wedge \nabla \delta(\vec{x}). \eeq

Using these expressions, it is straightforward to compute the static inter-vortex potential between two vortices with
orientations $\vec{n}_1$ and $\vec{n}_2$ at distance $R$:
\beq U=  2 \pi \left( c_0^2 K_0(e_0 \sqrt{\xi} R)
-b_1^2  K_0 (M_{S1} R) -b_2^2 K_0 (M_{S2} R) +\right. \eeq
\[ \left. +
(\vec{n}_1\cdot\vec{n}_2) ( c_3^2  K_0(e_3 \sqrt{\xi} R)
 -b_3^2  K_0 (M_{T1} R) -b_4^2  K_0 (M_{T2} R))
 \right). \]
In the BPS case ($\eta_0=\eta_3=0$) this potential is exactly zero, because
we have $M_{S1}=M_{S2}=e_0 \sqrt{\xi}$, $M_{T1}=M_{T2}=e_3 \sqrt{\xi}$
and $c_0^2=b_1^2+b_2^2$, $c_3^2=b_3^2+b_4^2$.

If $\eta_{3},\eta_0\neq 0$, we find that at large distance the particle with
lowest mass is the one which dominates the interaction.
We have always the following inequalities:
 \beq M_{S2} <  e_0 \sqrt{\xi} < M_{S1}, \,\,\,
M_{T2} <  e_3 \sqrt{\xi} < M_{T1}.
 \eeq
Thus if $M_{S2}<M_{T2}$ then we have:
 \beq U   \approx -2 \pi b_2^2 K_0 (M_{S2} R) \approx
  -2 \pi b_2^2 \sqrt{\frac{\pi}{2 M_{S2} R}} e^{-M_{S2} R }, \eeq
which gives an always  attractive force. On the other hand, if $M_{T2}<M_{S2}$:
 \beq U   \approx -2 \pi b_4^2 K_0 (M_{T2} R) (\vec{n}_1\cdot\vec{n}_2)  \approx
  -2 \pi b_4^2 \sqrt{\frac{\pi}{2 M_{T2} R}} e^{-M_{T2} R } (\vec{n}_1\cdot\vec{n}_2), \eeq
  which gives attraction for $\vec{n}_1=\vec{n}_2$ and
repulsion for $\vec{n}_1=-\vec{n}_2$.

A very peculiar thing happens for $e_0=e_3$ and $\eta_0=\eta_3\neq 0$.
For these fine-tuned
values of the couplings $M_{S2}=M_{T2}=M_2$ and $b_2=b_4$,
so the effective vortex potential has the form:
 \beq U   \approx -2 \pi b_2^2 K_0 (M_{2} R) (1+\vec{n}_1\cdot\vec{n}_2)  \approx
  -2 \pi b_2^2 \sqrt{\frac{\pi}{2 M_{2} R}} e^{-M_{2} R } (1+\vec{n}_1\cdot\vec{n}_2), \eeq
which gives a flat potential for $\vec{n}_1=-\vec{n}_2$.  This is consistent with the fact that in this limit
 the $(1,0)$ and the $(0,1)$
vortices do not interact because they are completely decoupled (see the argument below Eq.~(\ref{eq:11vor})). This
behaviour is similar to the one found in Ref.~\cite{verynewones} for global non-Abelian vortices.

\section{Effective worldsheet theory}

\subsection{Single vortex}

It is useful in the following to use the singular gauge in which the squarks fields at $r \rightarrow \infty $ tend to
a fixed VEV and do not wind. In this gauge, the ansatz (\ref{eq:single}) for the single vortex reduces to
\beq A^0_i=\frac{\epsilon_{ij} x_j}{r^2} f_0, \,\,\,
A^k_i=\frac{\epsilon_{ij} x_j}{r^2} f_3 n^k, \,\,\, a^0=\lambda_0, \,\,\,  a^k= n^k \lambda_3,\eeq
\[ Q=\frac{\phi_1  +\phi_2}{2} {\mathbf 1}+
\frac{\phi_1 -\phi_2}{2} \tau^k n^k. \]
We will assume that the orientational coordinates $\vec{n}$ are functions of the string worldsheet coordinates.
$\vec{n}$ becomes a field of a $1+1$ dimensional sigma model. This effective theory has no potential due to the fact
that the $n^k$ parameterize some zero modes; in the following we will compute the kinetic term. For the gauge field
components $A_{0,3}$, we will use the same ansatz as used in Refs.~\cite{nab,sy-cm}:
\beq A_k= -\frac{1}{2} (\tau^a \epsilon^{abc}
 n_b \partial_k n_c) \rho(r), \,\,\, k=0,3. \eeq
The field strength components $F_{ki}$ with $k=0,3$ and $i=1,2$ are not zero any more:
\beq F_{ki}=\frac12\,\de_k n^a \tau^a\epsilon_{ij}\,\frac{x_j}{r^2}\,
f_3[1-\rho(r)]+ \frac{1}{2} (\tau^a \epsilon^{abc}
 n_b \partial_k n_c)  \frac{x_i}{r}\frac{d}{dr}\rho(r).
\eeq

Substituting this expression into the kinetic term for the gauge field and for the squark fields, we obtain a simple
generalization of the BPS case discussed in Refs.~\cite{nab,sy-cm,Eto:2006uw}:
\beq S^{1+1}=\frac{\beta}{2} \int dt dz (\partial_j n^k)^2,\eeq
where:
\beq \beta=\frac{2 \pi}{e_3^2} \int r dr \left\{
\rho'^2+\frac{f_3^2 (1-\rho)^2}{r^2}+ \lambda_3^2 (1-\rho)^2+ \right. \label{bebel} \eeq
\[  \left.
+e_3^2 \{ (\phi_1^2+\phi_2^2)\frac{\rho^2}{2}+(1-\rho)(\phi_1-\phi_2)^2\} \right\}.
\]
We have to solve the Euler-Lagrange equations for $\rho(r)$, with $\rho(0)=1$ and $\rho(r\rightarrow\infty)=0$. In the
BPS case, where $\lambda_3$ is trivially $0$,
 we can show from the equations of motion~\cite{sy-cm} that
$\rho=1-\phi_1/\phi_2$ and that $\beta=2 \pi/ e_3^2$ (see also \cite{Eto:2006uw,Eto:2004rz}); in the general case
$\eta_3,\eta_0\neq 0$ there is not such powerful analytical result, here we have to solve the equations for $\rho$
numerically and then calculate $\beta$.

In the BPS case we have additional fermionic zero modes, associated with the unbroken supercharges;
 at small $\eta_0,\eta_3 \neq 0$ these modes should be still present, but they will not be
 described anymore by the fermionic sector of a supersymmetric effective theory in $1+1$ dimensions.
We will not discuss this aspect in this paper and we will leave it as a problem for further
investigation.

The color-flavor modes of the $(2,0)$ vortex are very similar to the $(1,0)$ ones: both the vortices have a
$\mathbb{CP}^1$ moduli space and the value of $\beta$ can be  determined using Eq.~(\ref{bebel}). For the $(1,1)$
vortex,nevertheless, these modes are just trivial because all the profile functions are proportional to the identity
matrix.

\subsection{Two well separated vortices}

A proper description of the system has to take into account also the quantum aspects of the sigma model physics. Let us
consider two vortices with internal orientations $\vec{n}_1,\vec{n}_2$. The relative distance between them can be
promoted to a complex field $R$; the global position of the system, on the other hand, decouples from the other the
degrees of freedom. If the distance of the two vortices is large ($|R|\rightarrow \infty$), we expect that the
effective worldvolume description of the bosonic degrees of freedom is:
\beq S= \int dt dz \left\{
 \frac{\beta}{2} (\partial_k n_1^a)^2+
\frac{\beta}{2} (\partial_k n_2^a)^2+ {T} |\partial_k R |^2 + v_s(|R|)+ v_t(|R|) \vec{n}_1\cdot\vec{n}_2 \right\} ,\eeq
where:
\beq v_s(|R|)=-2 \pi b_2^2 \sqrt{\frac{\pi}{2 M_{S2} |R|}}
 e^{-M_{S2} |R| },  \eeq
\[ v_t(|R|) =-2 \pi b_4^2 \sqrt{\frac{\pi}{2 M_{T2} |R|}} e^{-M_{T2} |R| }, \]
where $T$ is the tension of a single vortex. This description is  good only for large values of the VEV of the field
$R$; at $R=0$ the internal degrees of freedom are no longer described by $\mathbb{CP}^1 \times \mathbb{CP}^1$, but by a
space with topology $\mathbb{CP}^2/\mathbb{Z}_2$ (see Refs.~\cite{composite,EKMNOVY}). Moreover, the expression used
for the potential is good only for large vortex separation.

If we keep the VEV of $R$ fixed (which physically corresponds to keep the distance of the two vortices fixed with some
external device),
 the effective description is given by two  $\mathbb{CP}^1$ sigma models
 with a small interaction term of the form $c \, \vec{n}_1 \cdot \vec{n}_2$.

\section{Conclusion and Discussion}

For Abelian type I superconductors, the force between two vortices with the same winding number is always attractive.
This is true at large and at small distances, as shown by numerical calculations in Ref.~\cite{jr}. In the model
discussed in this paper, for $\eta_0,\eta_3 >0$,  the masses of some of the scalars fields are always found to be less
than the mass of the corresponding vector boson. In this sense we can think the system as a generalization of the
Abelian type I superconductor. However, here is an important difference: the force between two vortices is not always
attractive; there is a non-trivial dependence on the coupling, the relative internal orientation and the distance.

In this paper, we have studied the problem in two different limits: large vortices separation and coincident vortices.
For large separations we have computed the leading potential analytically; the behavior at large distance  is dominated
by the particle with the lowest mass $M_{\rm low}$. There are two main alternatives, which hold for different values of
the couplings:
\beq
U(R) \propto
\left\{
\begin{array}{cccl}
-   \sqrt{\frac{1}{2 M_{S2} R}}
 e^{-  M_{S2} R}  & {\rm for} & M_{low}=M_{S2},& {\rm Type} \, {\rm I}\\
 - (\vec{n}_1 \cdot \vec{n}_2)   \sqrt{\frac{1}{2 M_{T2} R}}
 e^{-  M_{T2} R}  & {\rm for} & M_{low}=M_{T2},& {\rm Type} \, {\rm I}^*
\end{array}
\right. \label{classif}
\eeq
where $M_{S2}$, $M_{T2}$ are the masses of the scalars in Eq.~(\ref{masse}). In order to distinguish these regimes, we
call them Type I and Type I$^*$; for Type I$^*$ vortices
 the sign of the asymptotic force depend on $\vec{n}_1 \cdot \vec{n}_2$.
For the fine-tuned values $e_0=e_3$ and $\eta_0=\eta_3\neq 0$,
 the relation $M_{S2}=M_{T2}=M_2$ holds,
and the effective vortex potential has the form:
 \beq U(R)  \propto
  - (1+\vec{n}_1\cdot\vec{n}_2)  \sqrt{\frac{1}{2 M_{2} R}} e^{-M_{2} R } , \eeq
which gives a flat potential for $\vec{n}_1=-\vec{n}_2$.

For coincident vortices we have found two stationary solutions of the equations of motion, the $(1,1)$ and the $(2,0)$
vortices, and we computed their tensions numerically. The results are shown in Fig.~\ref{titti}; both the cases
$T_{1,1}>T_{2,0}$ or $T_{2,0}>T_{1,1}$ are possible for different values of the coupling. The moduli space
interpolating between these solutions at $\eta_0=\eta_3=0$ disappears for non-zero values of one of these parameters
(see Fig.~\ref{confronto}).

It is interesting to match the data of the two complementary approaches. Let us for simplicity consider  the case of
parallel ($\vec{n}_1=\vec{n}_2$) and anti-parallel ($\vec{n}_1=-\vec{n}_2$)  vortices. In the case of parallel vortices
at large separation distance, the force is always attractive; also from numerical calculations we find $T_{2,0}<2
T_{1,0}$ for all the values of the coupling that we have analyzed.
 We have not made the calculation for arbitrary
 distances, but we think that the above are a good evidence for the fact
 that the force between two parallel vortices is always attractive in our model.

\begin{figure}[h]
\begin{center}
$\begin{array}{c@{\hspace{.2in}}c} \epsfxsize=2.5in
\epsffile{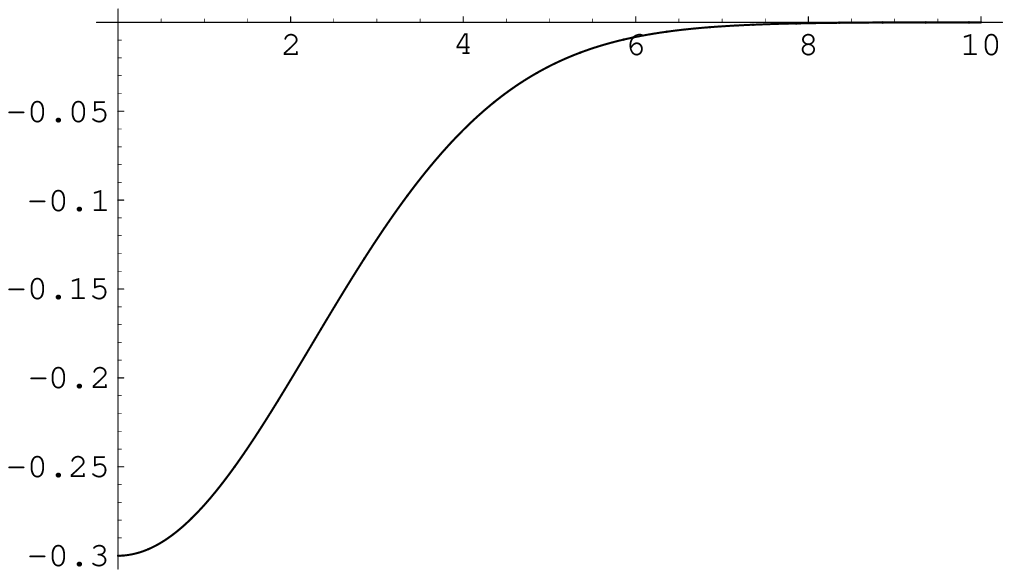} &
    \epsfxsize=2.5in
    \epsffile{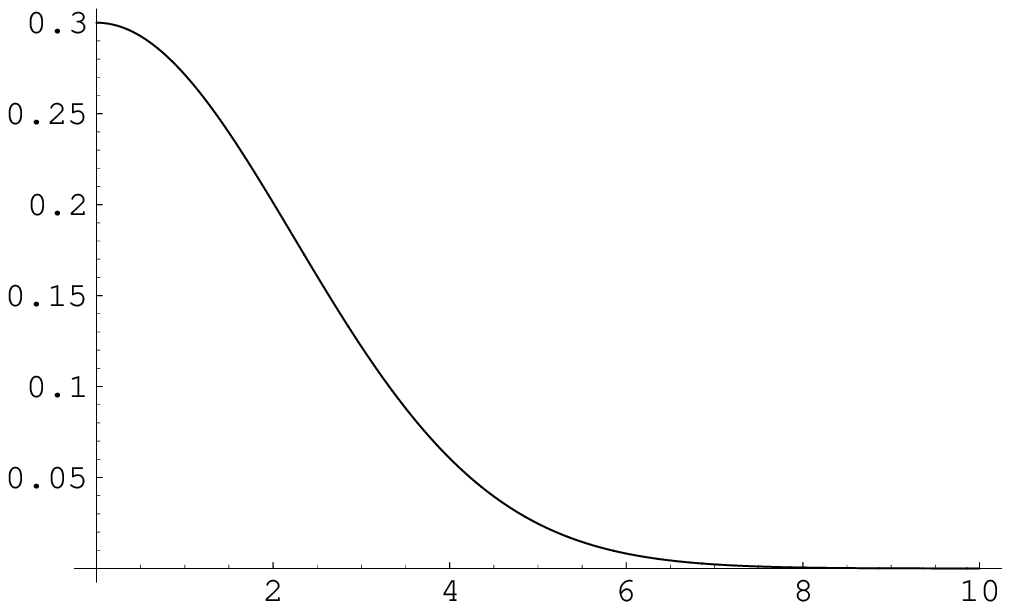}
\end{array}$
\end{center}
\caption{\footnotesize Qualitative plot of the vortex potential as function of the vortex distance
 for $\vec{n}_1=-\vec{n}_2$ and $e_0=e_3$.
For $\eta_0>\eta_3$ we have attraction (left); for $\eta_0<\eta_3$
we have repulsion (right). With the choice
$\eta_0=\eta_3$ there is no net classical force.
}
\label{qual0}
\end{figure}

For anti-parallel vortices, on the other hand, the situation is more complicated. At large distances there is
attraction if $M_{T2}>M_{S2}$ and repulsion if $M_{T2}<M_{S2}$. For the choice $e_0=e_3$ and $\eta_0=\eta_3$, the
relation $M_{T2}=M_{S2}$ holds; as we already noticed in Sect.~3, for these particular values the two diagonal $U(1)$
factors decouple: there is no net classical force  between the $(1,0)$ and the $(0,1)$ force for arbitrary distance
(this configuration is not stable, because if we allow $\vec{n}_1$, $\vec{n}_2$ to vary, we have that $(2,0)$ vortex
has a lower energy). Indeed, if we keep $e_0=e_3$, we obtain
 $M_{T2}>M_{S2}$, $T_{1,1}<2 T_{1,0}$ for $\eta_0>\eta_3$ and
$M_{T2}<M_{S2}$, $T_{1,1}>2 T_{1,0}$ for $\eta_0<\eta_3$. This is a good evidence that for  $\eta_0>\eta_3$ we have an
attractive force
 and for $\eta_0<\eta_3$ we have a repulsive one (see Fig.~(\ref{qual0})).

\begin{figure}[h]
\begin{center}
$\begin{array}{c@{\hspace{.2in}}c} \epsfxsize=2.5in
\epsffile{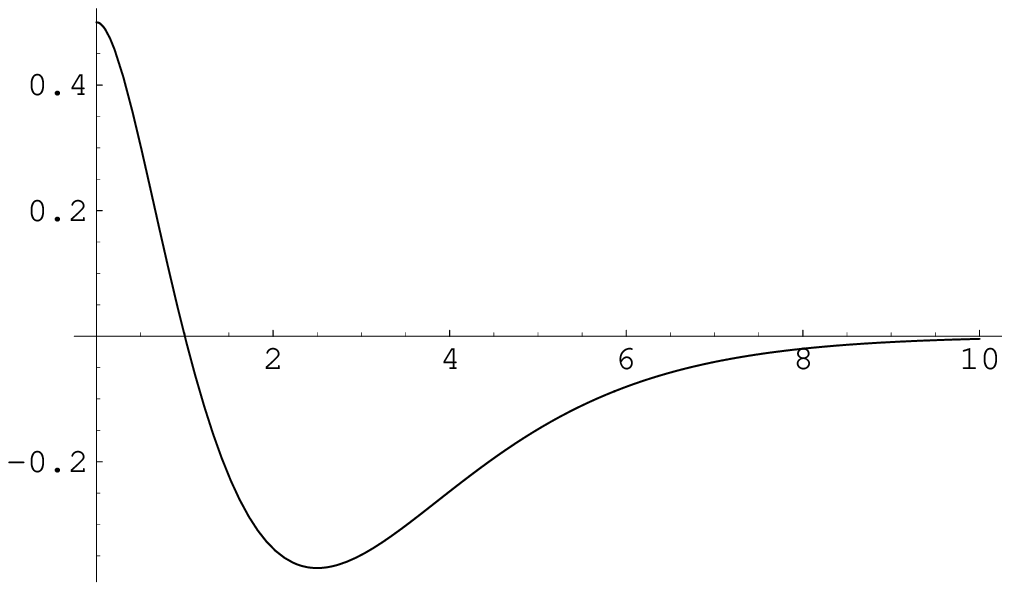} &
    \epsfxsize=2.5in
    \epsffile{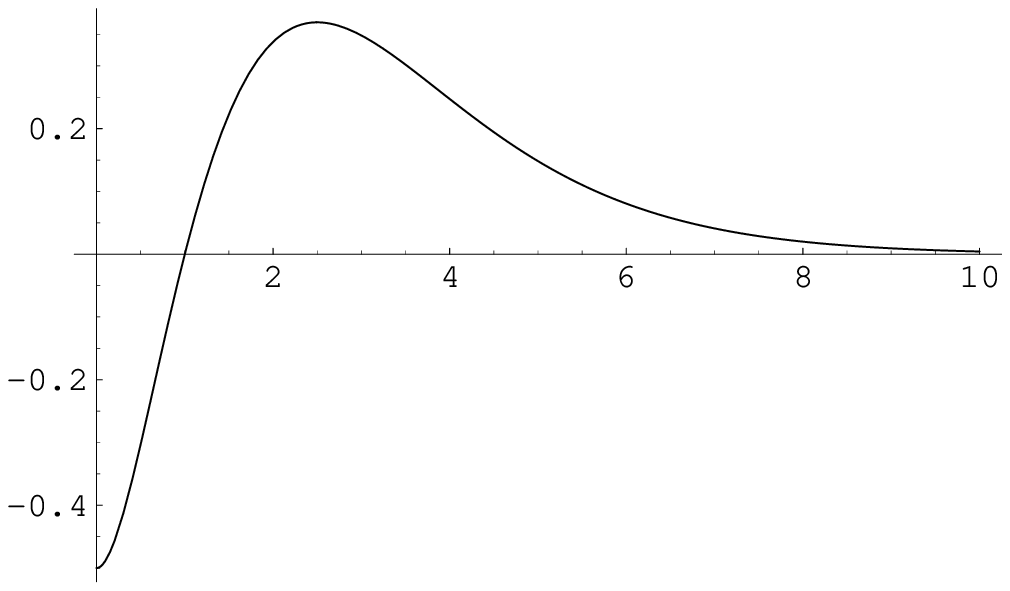}
\end{array}$
\end{center}
\caption{\footnotesize For $e_0\neq e_3$ the qualitative plot of the vortex potential as function of the vortex
distance
 for $\vec{n}_1=-\vec{n}_2$ can have maxima or minima for non-zero vortex separation.}
\label{qualitativo}
\end{figure}

If we relax the condition $e_0=e_3$, there are situations in which at large distance there is attraction (because
$M_{T2}>M_{S2}$) and also we get $T_{1,1}>2 T_{1,0}$, as is shown in  Fig.~\ref{qualitativo} to the left: this means
that there is a critical distance, in which there is a minimum of the inter-vortex potential for anti-parallel vortices
(if we allow the vortex orientation to flip, probably it will not be a minimum any more, because we still have
$T_{2,0}<2 T_{1,0}$). An example of this situation can be obtained with the couplings
$\xi=2,e_0=1/4,e_3=1/2,\eta_0=0,\eta_3=4$. Moreover we can obtain $M_{T2}<M_{S2}$ and $T_{1,1}<2 T_{1,0}$, which means
that there is a critical distance at which there is a maximum of the inter-vortex potential (see Fig.~\ref{qualitativo}
on the right). An example of this situation can be obtained with the couplings
$\xi=2,e_0=1/2,e_3=1/4,\eta_0=3.5,\eta_3=2$.

It is interesting that at $R=0$ there are two different regimes, which depend on the values of the coupling, with very
different properties. The physics of the $(2,0)$ vortex is described by a bosonic $\mathbb{CP}^1$ sigma model;  the
$(1,1)$ vortex on the other hand is an Abelian vortex with no internal degrees of freedom. For some values of the
coupling we have an evidence that both vortices are local minima of the tension (see Fig.~\ref{confronto}): one of the
two is metastable (indeed for some fine tuned values of $\eta_{0,3}$ we have that both the vortices have the same
tension).

A model with metastable vortices at weak coupling has already been studied in Ref.~\cite{metastable}. This behavior is
reminiscent of $SU(N)$ Yang-Mills, where for each topological  $n$-ality we can have different string tensions for each
representation of the Wilson Loop. In each topological sector there is just one stable string, corresponding to the
antisymmetric representation; there are evidences that the strings with  other representations are metastable strings,
at least in the large $N$ limit (see Ref.~\cite{as} for a discussion).

In this paper we have been interested in non-Abelian non-BPS vortices in an ${\cal N}=1$ supersymmetric model. In such a
restricted model we have found a physics similar to Abelian type I superconductors. In a companion paper, we will discuss a
simpler non-supersymmetric model in which we can have both type I and type II non-Abelian superconductivity.

\section*{Acknowledgments}

We are grateful to Toshiaki Fujimori, Bjarke Gudnason, Kenichi Konishi, Giacomo Marmorini, Muneto Nitta, Keisuke
Ohashi, Mikhail Shifman, David Tong and Alexei Yung for useful discussions and comments. The work of M.E.~is supported
by the Research Fellowships of the Japan Society for the Promotion of Science for Research Abroad.

\appendix

\section{Large $\eta_0$,$\eta_3$ limit}

If we take the limit $e_0 \eta_0 \gg \sqrt{\xi}$
 we can integrate out the superfield $a$
from the superpotential:
\beq a= \frac{  \xi - \Tr  \tilde{Q} Q}{2 \eta_0}. \eeq
Similarily way in the  limit $e_3 \eta_3 >> \sqrt{\xi}$ we can integrate out the superfield $a^k$:
\beq a^k=\frac{- \Tr \tilde{Q} \tau^k Q}{2 \eta_3}. \eeq
The effective superpotential is:
\beq W=-\frac{1}{\sqrt{2}} \left[
\frac{\Tr(\tilde{Q} \tau^k Q) \Tr(\tilde{Q} \tau^k Q)}{4 \eta_3}
+\frac{( \xi- \Tr(\tilde{Q} Q))^2}{4 \eta_0}\right]. \eeq
The potential is:
 \beqn
V&=& \frac{e_3^2}{8}
 \left(
 \Tr (Q^\dagger \tau^k Q) -\Tr(\tilde{Q} \tau^k \tilde{Q}^\dagger) \right)^2+
 \frac{e_0^2}{8} \left(\Tr (Q^\dagger Q)-\Tr(\tilde{Q} \tilde{Q}^\dagger) \right)^2+
 \nonumber\\[3mm]
  & +&  \left(  \frac{ | \xi - \Tr  \tilde{Q} Q|^2}{8 \eta_0^2}
  +\frac{|\Tr \tilde{Q} \tau^k Q|^2}{8 \eta_3^2}
  \right)
  \left( \Tr (Q^\dagger Q) + \Tr (\tilde{Q} \tilde{Q}^\dagger ) \right)+
 \nonumber\\[3mm]
  & +&  \frac{i \epsilon_{klm} \Tr (Q^\dagger \tau^k \tilde{Q}^\dagger)
  \Tr (\tilde{Q} \tau^l Q)
  }{8  \eta_3^2}
 \left( \Tr (Q^\dagger \tau^m Q) + \Tr (\tilde{Q} \tau^m \tilde{Q}^\dagger ) \right)-
 \nonumber\\[3mm]
  & -&  \frac{( \xi - \Tr  \tilde{Q}^\dagger Q^\dagger)  \Tr (\tilde{Q} \tau^k Q)
  + ( \xi - \Tr  \tilde{Q} Q) \Tr ( Q^\dagger \tau^k  \tilde{Q}^\dagger )
  }{8 \eta_0 \eta_3}
  \nonumber\\[3mm]
  & &
  \left( \Tr (Q^\dagger \tau^c Q) + \Tr (\tilde{Q} \tau^c \tilde{Q}^\dagger ) \right)
  \,.
 \eeqn
Note that in this low energy
action there is another vacuum at $Q=\tilde{Q}=0$.

 The equations of the $(p,k)$ vortices
  are a bit simpler, because we can integrate out
the adjoint fields $a$,$a_k$ and so we need less profile functions.
The energy is:
\beq \mathcal{E} = 2 \pi \int r dr \left(
\frac{f_0'^2}{2  e_0^2 r^2}+\frac{f_3'^2}{2  e_3^2 r^2}
+2(\phi_1'^2+\phi_2'^2)+
\right. \eeq
\[  \left. + \frac{(\phi_1^2+\phi_2^2)(f_0^2+f_3^2)+
2 f_3 f_0 (\phi_1^2-\phi_2^2) }{2 r^2}+\frac{(\phi_1^2+\phi_2^2)(\xi-\phi_1^2-\phi_2^2)^2}{4 \eta_0^2}+ \right.\]
\[ \left. +\frac{(\phi_1^2-\phi_2^2)^2  (\phi_1^2+\phi_2^2)}{4 \eta_3^2} - \frac{(\xi-\phi_1^2-\phi_2^2) (\phi_1^2-\phi_2^2)^2
}{2 \eta_0 \eta_3}\right). \]
The corresponding Euler-Lagrange equations are very similar to Eqs.~(\ref{2ordine}).

\begin{figure}[h]
\epsfxsize=7cm
\centerline{\epsfbox{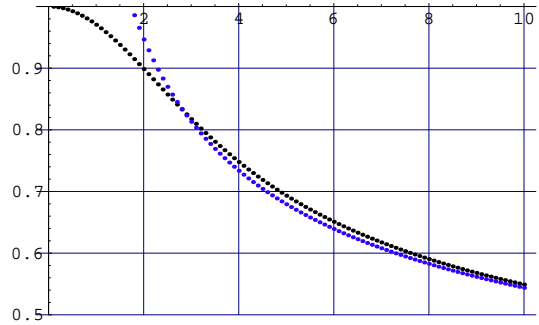}}
\caption{\footnotesize Tensions for the $(1,0)$ for $\eta_3=\eta_0=\eta$.
 The numerical values
 $e_0=1/2$, $e_3=1/4$, $\xi=2$, $0<\eta<10$ are used.
 The black dots give the result of the calculation in the
 full theory; the blue one give the result in the large $\eta_j$
 effective theory. For $\eta_j=0$ the vortex is of course BPS.}
\label{tensions}
\end{figure}

Numerical calculations can be performed for the profile functions and the tension. In Fig.~\ref{tensions} there is a
comparison between the tension calculated in the full theory and in the large $\eta_{0,3}$ approach. For small $\eta_j$
the correction in the tension from the BPS case is quadratic in $\eta_j$, as discussed in the Abelian case in Ref.
\cite{hou}.

The asymptotic profiles are also simpler in this limit.
Let us define:
\beq s=\delta \phi_1+ \delta \phi_2, \,\,\, d=\delta \phi_1- \delta \phi_2,
\eeq
we the find two linear differential equation:
\beq s''+\frac{s'}{r}-\frac{2 \xi^2}{\eta_0^2}s=0, \,\,\,
d''+\frac{d'}{r}-\frac{2 \xi^2}{\eta_3^2}d=0. \eeq
So the asymptotic solutions for the field profiles are:
\beq \delta \phi_{1,2} =
 s_0 \sqrt{\frac{1}{ r}} e^{-(\sqrt{2} \xi/\eta_0) r } \pm
 d_0 \sqrt{\frac{1}{ r}} e^{-(\sqrt{2} \xi/\eta_3) r }, \eeq
where $s_0,d_0$ are constants analogous to $b_k$.

\section{BPS equations}

In terms of fields, the BPS equations for two coincident vortices for $\eta_0,\eta_3=0$ read:
\beq F^a_{12}+e_3^2 \, \Tr (Q^{\dagger} \tau^a Q)=0, \,\,\,
 F^0_{12}+e_0^2 \, (\Tr (Q^{\dagger}  Q)-\xi)=0, \eeq
\[ (\nabla_1+i\nabla_2)\, Q=0\,.\]
In term of profiles functions the following system
of seven first order equations holds:
\beq \frac{f_0'}{r}=e_0^2 \left\{ \left( \cos \frac{\alpha}{2}\right)^2
(\kappa_1^2+\kappa_4^2)+ \left( \sin \frac{\alpha}{2}\right)^2
(\kappa_2^2+\kappa_3^2)-\xi\right\},
 \eeq
\[
\frac{f_3'}{r}=e_3^2 \left\{ \left( \cos \frac{\alpha}{2}\right)^2
(\kappa_1^2-\kappa_4^2)+ \left( \sin \frac{\alpha}{2}\right)^2
(\kappa_2^2-\kappa_3^2)\right\},
\]
\[ \frac{g'}{r}=e_3^2 \left\{ \kappa_1 \kappa_3-\kappa_2 \kappa_4 \right\},\]
\[\kappa_1'=  \frac{g-1}{r}  \sin ^2\left(\frac{\alpha }{2}\right) \kappa_3
+\frac{1-\cos (\alpha )+f_0+f_3}{2 r} \kappa_1,  \]
\[ \kappa _2'=-\frac{g-1}{r}  \cos ^2\left(\frac{\alpha }{2}\right)\kappa _4 -
\frac{1+\cos (\alpha )-f_0-f_3}{2 r} \kappa _2, \]
\[ \kappa _3'=\frac{g-1}{r}  \cos ^2\left(\frac{\alpha }{2}\right)\kappa _1 +
\frac{1+\cos (\alpha )+f_0-f_3}{2 r} \kappa _3, \]
\[ \kappa _4'=-\frac{g-1}{r}  \sin ^2\left(\frac{\alpha }{2}\right)\kappa _2 -
\frac{1-\cos (\alpha )-f_0+f_3}{2 r} \kappa _4. \]
In our numerical analysis of Sect.~6.1 we used the second order Euler-Lagrange equations;
at the end we used these first order equations as a check of our
calculation. We have found an excellent agreement between the two approaches.

\end{document}